\documentclass[fleqn,10pt]{wlscirep}
\title{Ergodicity breaking transition in a glassy soft sphere system at small but non-zero temperatures}

\author[1]{Moumita Maiti}
\author[1,*]{Michael Schmiedeberg}
\affil[1]{Institut f{\"u}r Theoretische Physik 1, Friedrich-Alexander-Universit{\"a}t Erlangen-N{\"u}rnberg, Staudtstr. 7, 91058 Erlangen, Germany}

\affil[*]{michael.schmiedeberg@fau.de}

\begin{abstract}
  While the glass transition at non-zero temperature seems to be hard to access for experimental, theoretical, or simulation studies, jamming at zero temperature has been studied in great detail. Motivated by the exploration of the energy landscape that has been successfully used to investigate athermal jamming, we introduce a new method that includes the possibility of the thermally excited crossing of energy barriers. We then determine whether the ground state configurations of a soft sphere system are accessible or not and as a consequence whether the system is ergodic or effectively non-ergodic. Interestingly, we find an transition where the system becomes effectively non-ergodic if the density is increased. The transition density in the limit of small but non-zero temperatures is independent of temperature and below the transition density of athermal jamming. This confirms recent computer simulation studies where athermal jamming occurs deep inside the glass phase. In addition, we show that the ergodicity breaking transition is in the universality class of directed percolation. Therefore, our approach not only makes the transition from an ergodic to an effectively non-ergodic systems easily accessible and helps to reveal its universality class but also shows that it is fundamentally different from athermal jamming.
\end{abstract}
\begin{document}
\flushbottom
\maketitle
Published in Scientific Reports {\bf 8}, 1837 (2018), www.nature.com/articles/s41598-018-20152-3
\thispagestyle{empty}

\noindent 

\section*{Introduction}

When increasing the density or decreasing the temperature many particulate systems reach a state where no longer any significant dynamics can be observed such that the system is in an amorphous, effectively solid state. Such a dramatic slowdown of the dynamics has
been observed in many systems \cite{Tabor,Angell,Debenedetti,Richardetal,HunterandWeeks}, even for particles with quite simple pair interactions as in the case of colloids \cite{HunterandWeeks,PuseyandMegen,PuseyandMegen1,kegel,weeks2000}. In simulations, a glassy slowdown even occurs in hard spheres systems \cite{WoodcockandAngell,Speedy,Heuer,Sear}. The origin of the associated glass transition as well as its fundamental properties are still under discussion (cf. \cite{Dyre,BerthierandBiroli}).

A solid amorphous state that on the first glimpse seems to be similar to a glass can be achieved by increasing the density at exactly zero temperature \cite{LubachevskyandStillinger,LiuandNagel1}. For example, in the protocol proposed by O'Hern et al. \cite{OhernLangerLiuNagel,OhernSilbertLiuNagel} one  starts with a random configuration of soft spheres that interact according to a finite-ranged repulsive interaction like a Hertzian or a harmonic potential. Then the local energy minimum is determined, i.e., the energy is minimized without crossing energy barriers. Note that the energy that is minimized is given as sum of all pair interaction energies. Depending on the packing fraction of the system, either all overlaps have been removed, which corresponds to a ground state and is called an unjammed system, or the configuration at the local minimum contains overlapping particles, which is called a jammed configuration. Note that such a jammed configuration obviously is not a ground state and that as a consequence jammed systems usually are not in equilibrium, because in principle ground states might still exist but are just not accessed. In the limit of large system size, jamming occurs at a well-defined packing fraction corresponding to random closed packing \cite{OhernLangerLiuNagel,OhernSilbertLiuNagel}. Note that starting with other configurations will lead to an athermal jamming transition with the same scaling behavior but a different transition packing fraction \cite{Chaudhuryetal,Bertrandetal}. In this article we explore how the athermal jamming transition changes if energy barriers are crossed due to thermal fluctuations during the quest to approach the ground state by energy minimization. We then study the transition from systems that reach a ground state to systems where the ground state is not accessible. We show that the onset of the effectively non-ergodic behavior is given by a directed percolation transition in time.

A unified jamming phase diagram \cite{LiuandNagel} has been proposed where athermal jamming is the endpoint of the glass transition line, i.e., of the jamming transition at non-zero temperature. Interestingly, recent theoretical studies \cite{krzakalaandKurchan,parisiandzamponi,JacquinBerthierandZamponi} and simulations \cite{zhangetal,BerthierandWitten,BerthierandWitten1,IkedaBerthierSollich,Wangetal} suggests that the athermal jamming transition might occur inside the glass phase at small but nonzero temperatures. 

In a recent work \cite{corwin} Morse and Corwin modified the athermal jamming protocols and force the particles to stay in contact during energy minimization. They then observe a percolation transition of clusters formed by locally rigid particles and relate it to the dynamical glass transition. To be specific, Morse and Corwin show for systems in three to six dimensions that the resulting rigidity transition occurs at packing fractions that are close to the packing fractions expected for the dynamical glass transition according to \cite{charb11}, e.g., in three dimensions they find a transition at the packing fraction $0.55(8)$ \cite{corwin} which is far below the athermal jamming packing fraction $\phi_J=0.639$ \cite{OhernSilbertLiuNagel} but close to the packing fraction of the dynamical glass transition at $0.571$ \cite{charb11}. Morse and Corwin argue that particles that are not locally rigid have more degrees of freedom to explore the configuration space than particles in large clusters of locally rigid particles \cite{corwin}. An open question remains, namely why should the particles stay in contact during the unjamming procedure? This question will be answered by our finding that due to rare thermal rearrangements the system can effectively be trapped in a region of the configuration space where a significant number of particles stays in contact. As a consequence, the ergodicity-breaking transition studied in this article is directly connected to the percolation transition described by Morse and Corwin.

\begin{figure}
\centerline{\includegraphics[width=0.77\linewidth]{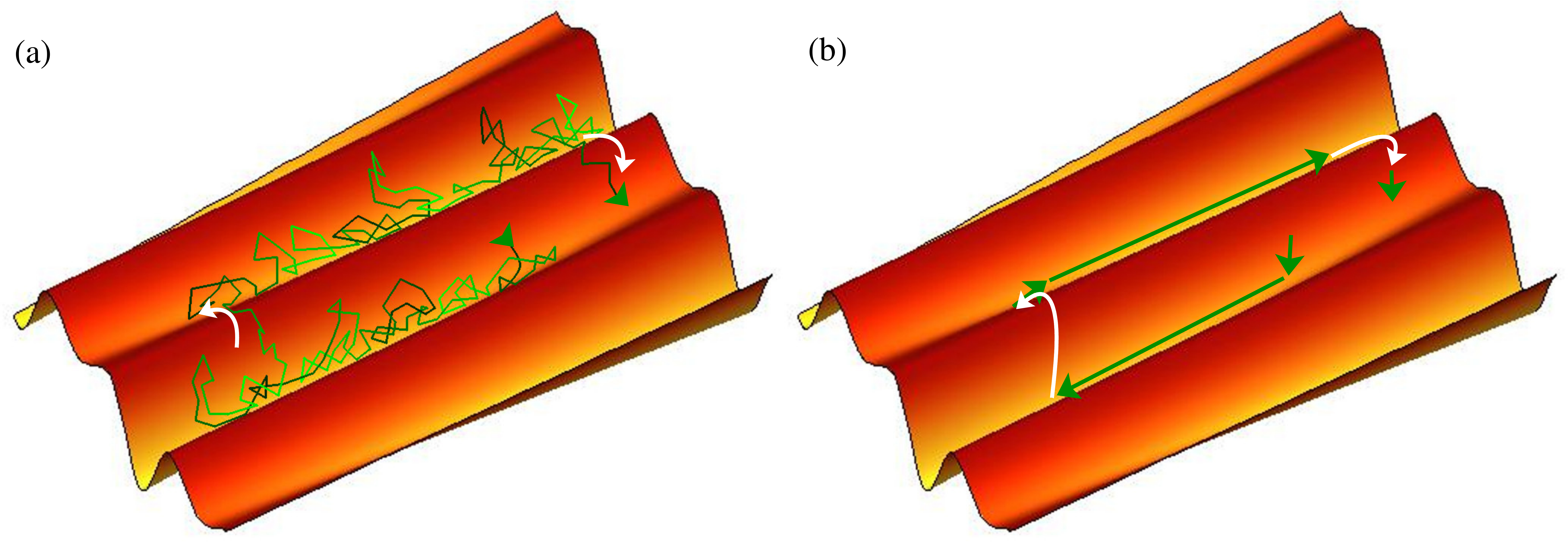}}
\caption{
  \sffamily \small {\bf {\sffamily Schematic of a trajectory.}}
  (a) Sketch of a trajectory in the $3N$-dimensional configuration space that mainly fluctuates in a valley of the energy landscape (green path), on average moves downhill, and rarely might also cross energy barriers (marked by white arrows). (b) In our approach we minimize the energy (green arrows) and with a given small probability $p$ cross energy barriers (white arrows). Note that the resulting ergodicity breaking transition for small $p$ does not depend on the choice of $p$. Fluctuations within a valley are not considered.}
\label{fig:model}
\end{figure}

{\bf Model: Testing the accessibility of ground states.}
In this work we study the competition and interplay of the slow relaxation by energy minimization within an energy basin and the rare crossing of barriers. The minimization part of our protocol is known from athermal jamming \cite{OhernLangerLiuNagel,OhernSilbertLiuNagel}, the hopping over barriers is often employed in models that are used in order to describe dynamical properties of glasses, e.g., dynamics heterogeneities or ageing \cite{Debenedetti,buechner00,denny03,doliwa03}. Note that in the later models usually an energy landscape of disconnected basins is assumed where thermalization within a basin takes place almost instantaneously. However for the case that we are interested in, i.e., for densities below athermal jamming, all energy basins are connected in case of an infinite system. The dynamics still can be slow in case of a slow relaxation within this one basin due to narrow and long pathways to the ground state.

If we explored the energy landscape by a Brownian dynamics simulation or a local Monte Carlo simulation, the trajectory in configuration space mainly would fluctuate in a valley and on average goes downhill as far as possible (see sketch in Fig. \ref{fig:model}(a)). On rare occasions an energy barrier can be crossed. Since we want to study large systems and long timescales, we consider a simplified approach in order to determine which parts of the configuration space can be accessed within a reasonable time. We start with random configurations of systems consisting of monodisperse spheres in three dimensions and then usually employ the energy minimization protocol of athermal jamming. However, we introduce additional steps that for each particle and each minimization step can occur with a small probability $p$ and can result in the crossing of an energy barrier (see sketch in Fig. \ref{fig:model}(b)). We have tested different implementations of the minimization as well as the barrier hopping steps. Details of the protocol that we usually use in the main text are given in the methods section and alternative protocols are discussed in the Supplementary Notes 1. All of these protocols lead to the same ergodicity-breaking transition in the limit of small $p$. Note that close to the observed transition we employ large systems of up to $N=10^7$ particles in order to avoid finite size effects that are explored in the Supplementary Notes 2.

The goal of the work is to find out whether the ground state can be reached or not. In analogy to the terminonoly used for athermal jamming, a system is considered to be unjammed if all overlaps can be eliminated. If for a larger packing fraction the number of particles that still possess overlaps is not decreasing, the system is called jammed. Note that we are only interested in the case of rare barrier crossing events, i.e., small $p$-values, where the dynamics is dominated by the minimization process and indeed effectively is stuck, if the system cannot reach the ground state. For large $p$ a significant number of rearrangement occurs due to the barrier crossing events such that instead of a dynamically jammed system one finds a fluid of soft, overlapping spheres. If all steps were random (for $p=1$) our protocol is similar to the one with only random displacements that we studied in \cite{MilzandSchmiedeberg}. The transition in this case was shown to be in the universality class of directed percolation and can be mapped onto a random organization transition, which was first observed in cyclically sheared colloidal systems \cite{Pine,Corte}.

In this article we identify the barrier crossing probability $p$ with temperature $T$ in the sense that for $T=0$ no barriers can be crossed, i.e., $p=0$, that for $T>0$ there is a non-zero probability $p>0$ of barrier crossing, and that with increasing temperature the probability to cross barriers increases as well. We want to point out that in order to obtain an ensemble sampling at a fixed real temperature $T$ of the configurations that we find, one would have to weight the observed trajectories by appropriate factors, i.e., by using Kramers' rate \cite{Kramers} for each barrier crossing. Note that in the limit of small $p$ that we are interested in only a minority of the particles cross a barrier at all. Therefore, in principle, an ensemble sampling based on our method is possible. Unfortunatally close to the glass transition such a sampling is computationally too expensive because large systems are required to avoid finite size effects. Note that Brownian dynamics or Monte Carlo simulations close to the glass transition also are too demanding if you wanted to study the critical behavior in the same system size as we do.
Since our main interest in this paper is to find out, whether the ground state is accessible or not, ensemble sampling or the determination of statistical, non-zero weights for accessible configurations is not necessary, because the accessibility of a configuration would not change if it was weighted differently.
Furthermore, any sampling of the power laws that we observe in the following would again lead to power laws with the same exponents.
Finally, we show that the transition density that we find does not depend on $p$ in the limit of small $p$.
The reason is related to a property of the configuration space, namely a spatial percolation transition that occurs at this density as we will show in the paragraph on spatial percolation and that also has been reported in \cite{corwin}. If a particle crosses an energy barrier it can affect the minimization process of all particles that are part of a percolated chain of overlapping particles. As a consequence the naive expectation that p directly determines how many particles are disturbed on their way towards a non-overlapping ground state turns out to be wrong in a spatially percolated system. In such a system the whole percolated chain of particles is affected if only one particle of the chain crosses an energy barrier. Therefore, the ergodicity breaking transition then does not depend on $p$ but is given by the packing fraction of the spatial percolation transition. 
Note that since $p$ is given by a strongly monotonic function of $T$, the ergodicity breaking transition also cannot depend on $T$ for small $T$ no matter how the functional dependence of $p$ on $T$ actually looks like.

\section*{Results}
Our new method for $p=0$, i.e., without any 
crossing of energy barriers, leads to the well-known athermal jamming transition at a packing fraction of $\phi_{\textnormal{J}}= 0.638$ which is in agreement with the results for a monodisperse system reported in \cite{OhernLangerLiuNagel,OhernSilbertLiuNagel}. 
As we will show in the following, for $p>0$ a different type of transition occurs which we call the thermal jamming transition.

\begin{figure}
\centerline{\includegraphics[width=0.65\linewidth]{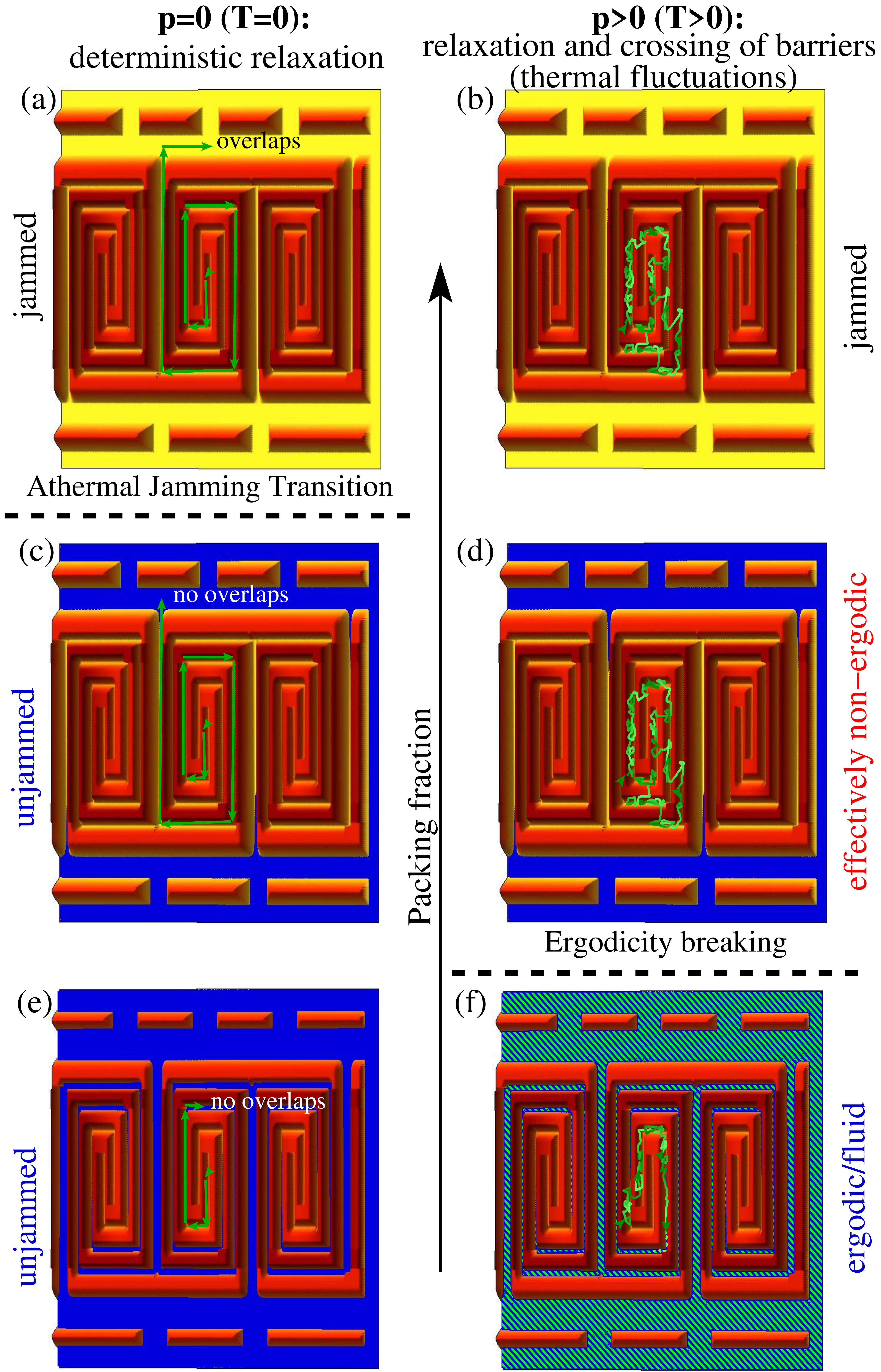}}
\caption{
  \sffamily \small {\bf {\sffamily Schematic of the energy landscape.}}
  Schematics of an energy landscape given by the sum of all pair interaction energies 
in configuration space spanned by the coordinates of all particles
at different densities and for different protocols. 
Note that below athermal jamming the energy basins are connected.
(a,c,e) Protocol leading to athermal jamming which is based on deterministic minimization steps (green arrows). (b,d,f) 
Exploration of configuration space in case of nonzero temperature where the rare crossing of energy barriers is possible. The blue area denotes zero energy states, i.e., unjammed configurations. (a,b) depict the case of a large packing fraction, (c,d) of an intermediate, and (e,f) of a small one. While the athermal jamming transition occurs at a packing fraction where the local minimum no longer is a zero energy state, in case of thermal jamming the system might fail to remove all overlaps even at a lower packing fraction, e.g. in (d), because it does not reach the unjammed local energy minimum. 
However, if unjammed configurations are reached, e.g., in (f), the system can explore easily the unjammed part of the configuration space as denoted by the green-blue hatched area in (f).
}
\label{fig:fig1}
\end{figure}

{\bf The thermal jamming transition.} 
To get a rough idea of the athermal and the thermal relaxation process, we sketch schematic energy landscapes in Fig.\ \ref{fig:fig1}. The blue areas correspond to unjammed ground states. In the dilute systems shown in Figs.\ \ref{fig:fig1} (e,f)  a lot of unjammed configurations occur while for large packing fractions 
(cf. Figs.\ \ref{fig:fig1} (a,b))
overlaps might prevail. In the energy landscapes on the left hand side of Fig.\ \ref{fig:fig1} the relaxation process occurs according to the athermal jamming protocol ($p=0$) where the local energy minimum is finally reached. The athermal jamming transition takes place at the packing fraction where the energy of the local minimum changes from zero to a non-zero value.

In case of thermal jamming ($p>0$) depicted on the right hand side of Fig.\ \ref{fig:fig1} 
the crossing of barriers is possible leading to a transition packing fraction that is below the one of athermal jamming.

 \begin{figure}
\centering
\includegraphics[width=\linewidth]{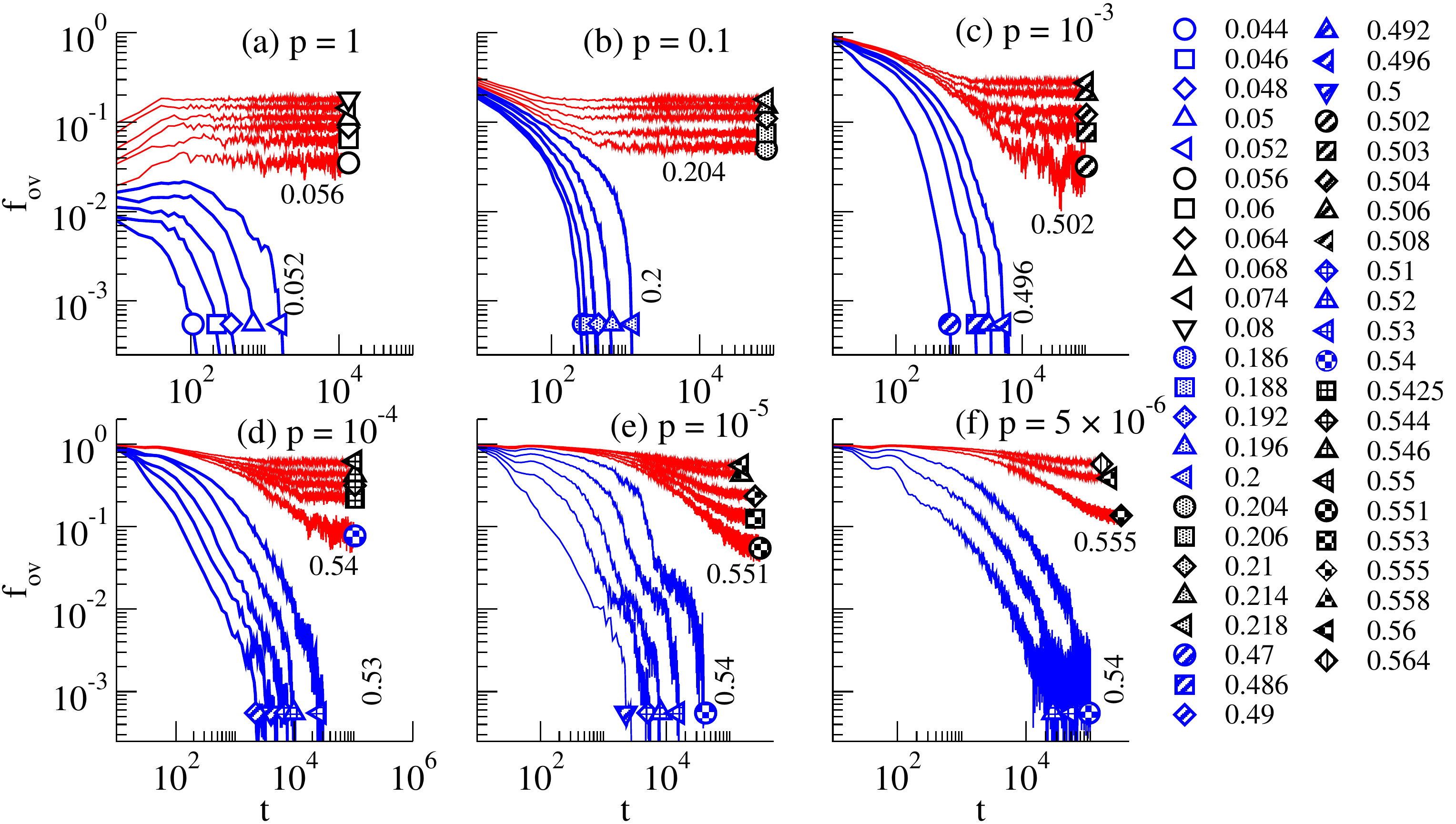}
\caption{
  \sffamily \small {\bf {\sffamily Transition between overlapping and non-overlapping final configurations.}}
  Fraction of overlapping particles $f_{ov}=N_{\textnormal{ov}}/N$ as a function of simulation steps $t$ for (a) $p = 1$, (b)$p = 0.1$, (c) $p = 0.001$, (d) $p = 10^{-4}$, (e) $p = 10^{-5}$, and (f) $p = 5\cdot 10^{-6}$. The curves below the transition (plotted in blue) relax to the zero energy state, while above the transition (red) they approach a steady state. The symbols denote the different densities and for each $p$ the packing fraction right below and right above the transition are given in addition in the figures.}
\label{fig:fig2}
\end{figure}

Fig.~\ref{fig:fig2} shows the number of overlapping particles as a function of the number of steps $t$ for selected $p$ and different $\phi$. For a given $p$ one finds unjammed configurations where the number of overlaps decays to zero at small packing fractions. At large packing fractions, the curves reach a steady state denoting a jammed system. Pair distribution functions $g(r)$ of jammed configurations close to the transition are analyzed in the supplementary note 3. We observe a pronounced peak of $g(r)$ close to $r=\sigma$, which is a known feature of soft sphere glasses close to the hard sphere limit \cite{parisiandzamponi,zhangetal}.

If 
one tries to cross barriers in all steps, i.e., for $p=1$ depicted in Fig.~\ref{fig:fig2}(a),
we observe a transition between $\phi = 0.052$ and $\phi = 0.056$. This case 
is similar to the transition studied in
\cite{MilzandSchmiedeberg,Frenkel} 
which has the same universality class as the random organization transition considered in
\cite{Pine,Corte,TjhungandBerthier}. 
Note that it is known that
 differences in the details of the protocol 
lead to different transition packing fractions. It was shown that the critical behavior either corresponds to a directed percolation transition either with a conserved number of binding sites particles or to one with an unconserved number \cite{MilzandSchmiedeberg,Corte}. Whether it is a non-conserved or a conserved directed percolation turned out to be hard to determine from an analysis of the critical exponents \cite{MilzandSchmiedeberg,Corte}. Note that binding sites correspond to overlapping particles in our case. Since there is no reason why the number of overlapping particles should be conserved and since a detailed analysis
in \cite{MilzandSchmiedeberg}
of the random organization transition 
indicates that it is more likely to be a non-conserved directed percolation transition,
we will compare our observed critical behavior to non-conserved directed percolation in the following.

If the probability $p$ is decreased, the observed transition packing fractions usually increases. However, in all cases with $p\leq 10^{-4}$ (Fig.~\ref{fig:fig2}(d-f)) the transition occurs at roughly the same packing fraction between $\phi = 0.53$ and $\phi = 0.555$. Furthermore, this transition packing fraction is much smaller than the athermal jamming transition packing fraction $\phi_{\textnormal{J}}= 0.638$ which is obtained for $p=0$.

\begin{figure}
\centering
\includegraphics[width=0.7\linewidth]{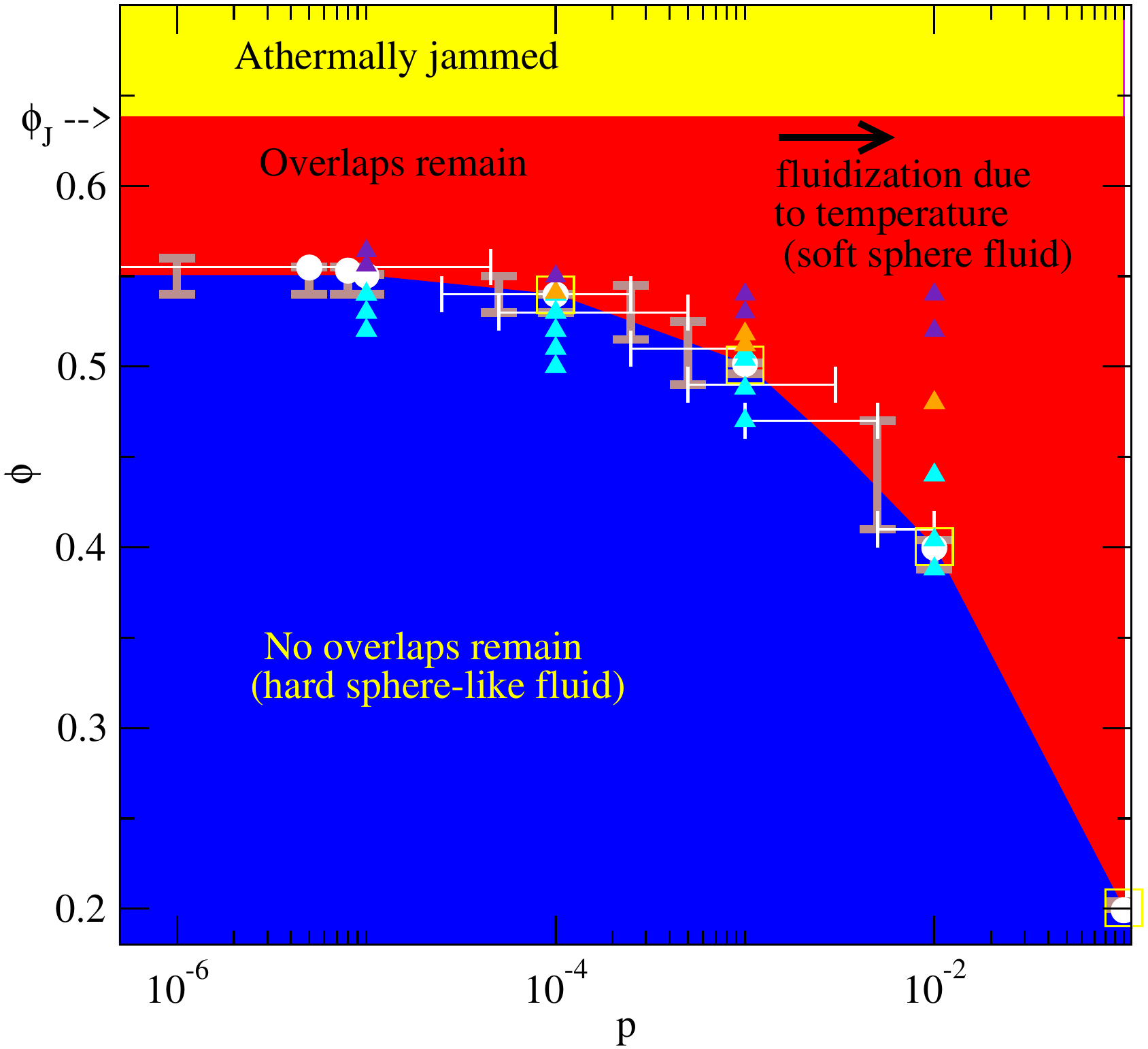}
\caption{
  \sffamily \small {\bf {\sffamily Thermal jamming phase diagram.}}
  Thermal jamming phase diagram showing the transition between 
states that can lead to non-overlapping configurations (blue) and states where overlaps remain (red)
depending on the probability $p$ of random steps (corresponding to a temperature $T$) and the packing fraction $\phi$. 
Note that we are especially interested in the case of small $p$ where the blue area denotes ergodic, unjammed states while the red area corresponds to thermally jammed states that are effectively non-ergodic.
The yellow area marks the packing fractions where the system would be athermally jammed in case of $p=0$. The white and brown bars denote the range in which the transition occurs in case $\phi$ or $p$ is kept constant, respectively. 
In addition we show  the transition packing fractions that we obtain from fits of critical power laws to the steady state values $f_{\textnormal{ov}}(t \rightarrow \infty)$ in case of jammed states (white solid circles, cf. Fig.~\ref{fig:fig4}(a)) or to the relaxation times $\tau$ (yellow open squares, cf. Fig.~\ref{fig:fig4}(b)). Finally, the triangles indicate where we explore whether chains of touching or overlapping particles in the final configurations are unpercolated (cyan triangles), continuously percolated in space (orange triangles), or spatially directed percolated (blue triangles). These spatial percolation transitions are analyzed in Fig.~\ref{fig:fig5}.}
\label{fig:fig3}
\end{figure} 

{\bf Thermal jamming phase diagram.} 
In Fig.~\ref{fig:fig3} we show how the 
transition packing fraction 
that separates states leading to non-overlapping configurations from states where overlaps remain
depends on the probability $p$. 
All overlaps can be removed in the blue area while in the red area the overlaps do not vanish.
For comparison we also depict where in the case $p=0$ the transition into an athermally jammed state occurs which is marked in yellow. The transition line 
is determined by the 
largest $\phi$ denoting a
state 
without remaining overlaps
the smallest $\phi$
of all states
that possess overlaps in the end
for a given $p$ (marked by brown error bars in Fig.~\ref{fig:fig3}). Similarly, for given $\phi$ the transition ranges in $p$ are determined (white error bars). In addition we obtained transition packing fractions by analyzing the critical behavior (white solid circles and yellow open square) as we will explain later. We have checked that the shown results are not affected by system size effects (see also Supplemetary Note 2.).

We find that the transition line $\phi_{\textnormal{c}}(p)$ of 
this transition that we term the thermal jamming transition
for $p\rightarrow 0$ approaches a packing fraction $\phi_{\textnormal{G}}=\lim_{p\rightarrow 0}\phi_{\textnormal{c}}(p)=0.55\pm 0.01$ where it denotes a transition between an ergodic state and an effectively non-ergodic state as we have explained before. The packing fraction $\phi_{\textnormal{G}}$ of this ergodicity breaking
in the limit of small but non-zero $p$ corresponding to small but non-zero temperatures significantly differs from the athermal jamming packing fraction $\phi_{\textnormal{J}}$. Note that if $p$ is not small the thermal rearrangements lead to a fluidization even in the case where overlaps remain. Therefore, the glass that we are interested in only occurs in the limit of small $p$.

\begin{figure}
\centering
\includegraphics[width=\linewidth]{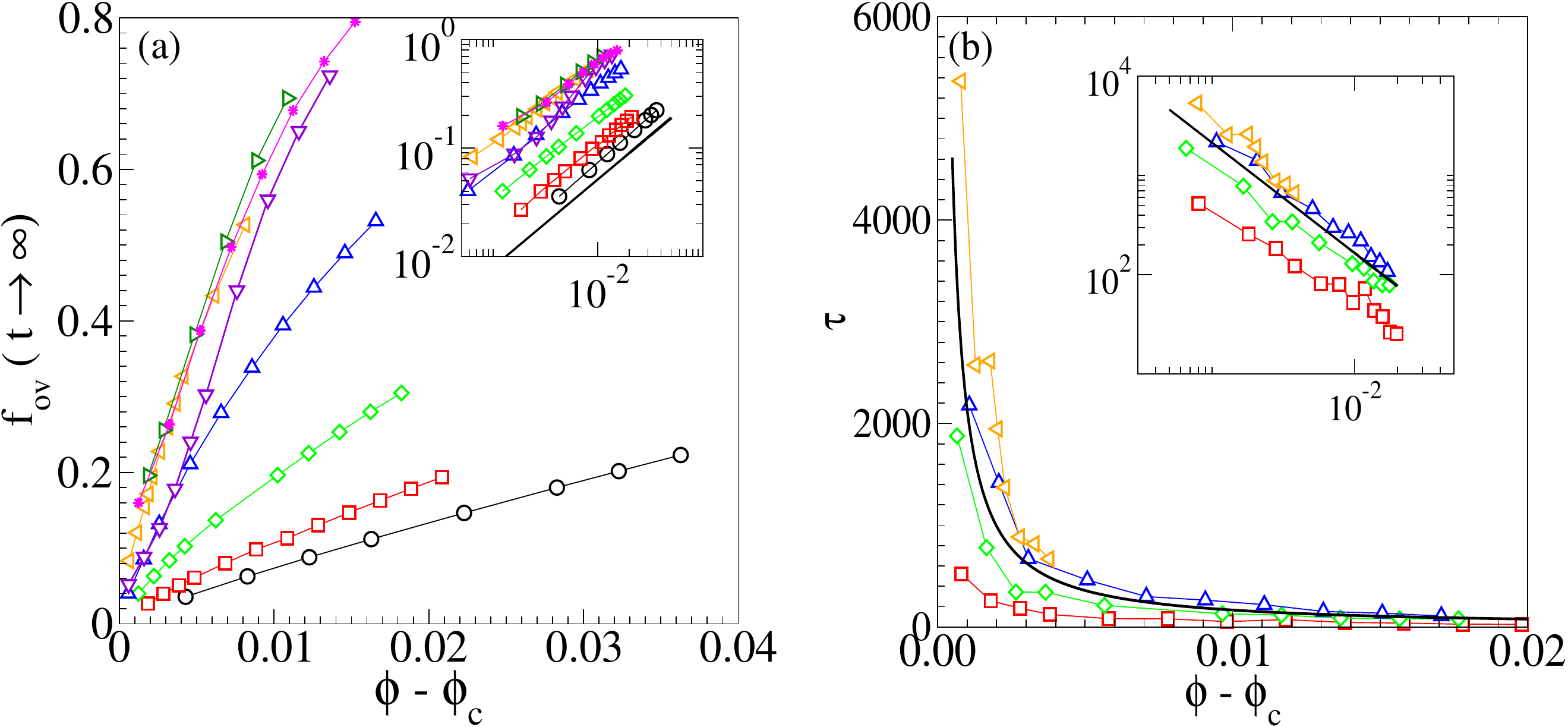}
\caption{
  \sffamily \small {\bf {\sffamily Critical behavior.}}
  Critical scaling close to the thermal jamming transition. (a) Fraction of overlapping particles in the long-time limit $f_{ov}(t \rightarrow \infty)$ and  (b) relaxation time $\tau$ as functions of the packing fraction $\phi$ minus the transition packing fraction $\phi_{\textnormal{c}}(p)$. 
Note that $\tau$ denotes the time at which the steady state is approached and not the time for the relaxation towards equilibrium.
Different probabilities $p$ are considered:: $p = 1$ (black), $p=0.1$ (red), $p=0.01$ (green), $p=0.001$ (blue), $p=0.0001$ (orange), $p=10^{-5}$(violet), $p=8 \times 10^{-6}$(dark green), and $p=5 \times 10^{-6}$ (magenta). The insets shows the same plots in log-log scale. The black lines indicate the exponents of power laws that occur in case of a directed percolation transition \cite{Hinrichsen}, i.e., exponent $\beta = 0.81$ in (a), and $\nu = -1.1$ in (b).}
\label{fig:fig4}
\end{figure}

{\bf Critical behavior.} 
Close to the thermal jamming transition we analyze the critical behavior. We determine the fraction of overlapping particles in the long-time limit $f_{\textnormal{ov}}(t \rightarrow \infty)$ as well as the relaxation time $\tau$ 
for the relaxation towards the state that we find for long times
as functions of $\phi - \phi_{\textnormal{c}}$.

In order to determine $f_{\textnormal{ov}}(t \rightarrow \infty)$ and $\tau$ we fit
\begin{equation}
\label{eq:fov}
f_{\textnormal{ov}}(t) = f_{\textnormal{ov}}(t \rightarrow \infty) + At^{\alpha}\times \exp(-\frac{t}{\tau}) 
\end{equation}      
to the relaxation curves above the jamming transition. This approach assumes a critical power law $f_{\textnormal{ov}}(t)\propto t^{\alpha}$ at the transition density from which the curves at other densities deviate at a time $\tau$. Such an approach already has been used in \cite{ReichhardtandReichhardt,MilzandSchmiedeberg}. As shown in Supplementary Note 3, our results are in agreement with $\alpha = -0.732$ as expected for directed percolation \cite{Hinrichsen}. For the fits to Eq.\ (\ref{eq:fov}), we use a fixed $\alpha = -0.732$ and otherwise employ $f_{\textnormal{ov}}(t \rightarrow \infty)$, the prefactor $A$, and the relaxation time $\tau$ as fitting parameters. For $\phi<\phi_{c}$ we often observe a power law decay with exponent $-1.5$ (see Supplementary Note 2), which does not influence the transition but makes it hard to define a relaxation time $\tau$ below $\phi_{\textnormal{c}}$.

In Fig.~\ref{fig:fig4}(a) $f_{\textnormal{ov}}(t \rightarrow \infty)$ is shown as function of $\phi - \phi_{\textnormal{c}}$. It is zero for unjammed configurations. Above the thermal jamming transition at $\phi_{\textnormal{c}}$ we find that our results can be described by a power law
\begin{equation}
\label{eq:beta}
f_{\textnormal{ov}}(t \rightarrow \infty) \sim (\phi - \phi_{\textnormal{c}})^{\beta}
\end{equation}      
with a critical exponent $\beta$. As can be seen in the log-log representation shown in the inset of Fig.~\ref{fig:fig4}(a) the exponent $\beta$ does not depend on $p$ and therefore all curves possess the same critical behavior as 
directed percolation. For comparison, the black line indicates the slope $\beta = 0.81$ expected for a directed percolation transition \cite{Hinrichsen}. The packing fractions that are obtained by fitting the power law in Eq.\ (\ref{eq:beta}) to our simulation data are shown with white solid circles in Fig.\ \ref{fig:fig3}. 

In Fig.~\ref{fig:fig4}(b) we show that $\tau$ as a function of $\phi - \phi_{\textnormal{c}}$ obeys a critical power law of the form
\begin{equation}
\label{eq:nu}
\tau \sim (\phi - \phi_{\textnormal{c}}^{\tau})^{\nu}
\end{equation}      
with an exponent $\nu$. The log-log plot in the inset of Fig.~\ref{fig:fig4}(b) demonstrates that $\nu$ also does not depend on $p$ and that all simulation results are in agreement with $\nu = -1.1$ (black line) as expected for directed percolation \cite{Hinrichsen}. The packing fractions where we observe the divergence of fitted power laws are shown with yellow open squares in Fig.\ \ref{fig:fig3}. 

In conclusions, the critical behavior for all $p>0$ is the same as 
directed percolation. Physically this result can be motivated in the following way: A system is thermally jammed if there is a region of overlapping particles that does never disappear as time proceeds though it might move around. This corresponds to a path that describes the propagation of overlaps in time and that is directed percolated (directed because the time is always directed).

The athermal jamming transition is not in the universality class of directed percolation. In fact, it is a very different transition because the number of overlaps per particles jump from $0$ to the value needed for isostaticity (e.g., to $4$ in two dimensions and $6$ in three dimensions) \cite{OhernLangerLiuNagel,OhernSilbertLiuNagel}. Therefore, even without taking the transition packing fractions into account one can conclude from the critical behavior that the athermal jamming transition cannot be the $p\rightarrow 0$ (or $T\rightarrow 0$) limit of the thermal jamming transition. Note that also the pair correlation function $g(r)$ differs significantly in thermally jammed and athermally jammed packings as we show in Supplementary Note 3.

\begin{figure}
\centering
\includegraphics[width=1\linewidth]{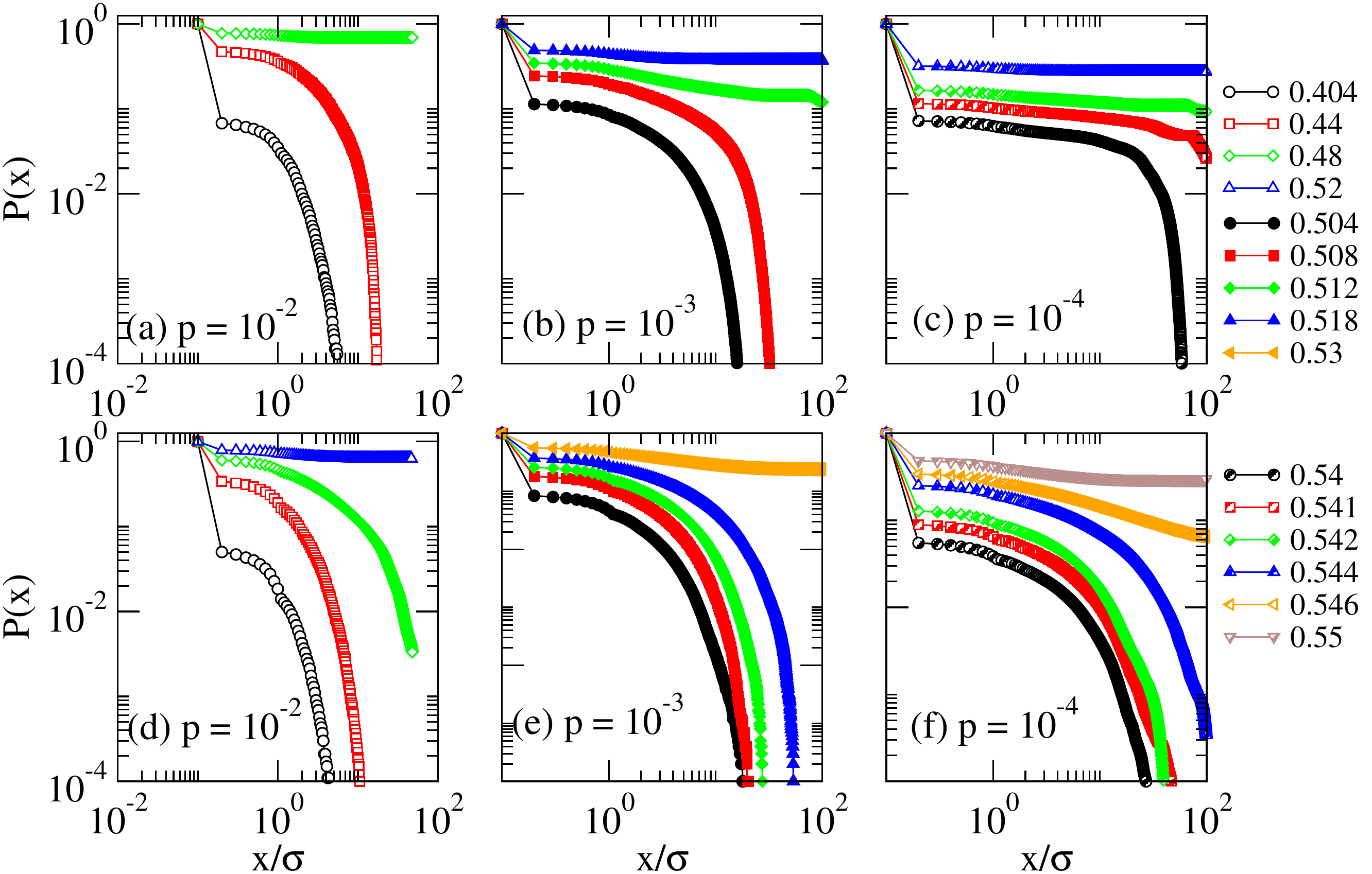}
\caption{
  \sffamily \small {\bf {\sffamily Spatial percolation transition.}}
  Probability distribution of $P(x)$ that there is a path within a cluster of connected particles that reaches longer than $x$ in $x$-direction. In (a), (b), and (c) all possible paths are considered such that a continuous percolation in space is obtained if $P(x)$ does not decay while the system is not percolated if $P(x)$ decays. In (d), (e), and (f) only directed paths in $x$-direction are considered indicating a directed percolated structure in space if $P(x)$ does not decay. The colors indicate the packing fractions and the probabilities $p$ are given in the figures. The unit of length is given by the diameter $\sigma$ of the particles.}
\label{fig:fig5}
\end{figure}

{\bf Spatial Percolation.} 
\label{spatial}
We analyze whether a spatial percolation transition occurs for the configurations obtained by the thermal jamming protocol.  We consider two particles to be in contact if they touch or overlap. Starting at an arbitrary particle we determine the cluster of particles that can be connected by contacts. In Fig.~\ref{fig:fig5}(a,b,c) we show for $p = 10^{-2}$, $10^{-3}$, and $10^{-4}$ and various $\phi$ the probability distribution $P(x)$ that a particle  still is in this contact cluster if in $x$-direction (or any other given direction) it is at a distance $x$ from the starting particle. For large $\phi$ we observe that the cluster of connected particles reaches through the whole system and therefore there is a continuous percolation transition in space. If we only consider paths within the clusters that are directed in $x$-direction, i.e., if we only go from particle to particle within the cluster if this increases the $x$-coordinate, we obtain a directed percolation transition in space as is shown in Fig.~\ref{fig:fig5}(d,e,f).

For the three values of $p$, we mark the studied state points in Fig.~\ref{fig:fig3} with triangles. Spatially unpercolated configurations are denoted by cyan triangles, continuously percolated states by orange triangles, and configurations that are also directed percolated in space by blue triangles. For large $p$ continuous and directed percolation take place deep in the thermally jammed phase. However, for small $p$ the percolation transitions both occur at a similar packing fraction as the thermal jamming transition. Therefore, our results indicates that one of these spatial percolation transition might be the reason why for small $p$ there is no significant increase of the thermal jamming transition packing fraction $\phi_{\textnormal{c}}(p)$ upon a decrease of $p$. As a consequence of the spatial percolation a single randomly moved particle can affect the whole system no matter how large it is and therefore even in the limit $p\rightarrow 0$ the relaxation into an unjammed state can be prevented. 

Morse and Corwin in their work \cite{corwin} find a spatial percolation of locally rigid particles at the same packing fraction up to the precision of our simulations. They claim that this transition is an echo of the dynamical glass transition \cite{corwin}. As we point out, it is related to the breaking of the effective ergodicity.

{\bf Packing fraction of ergodicity breaking for different starting configurations.} 
\label{sec:startcond}
We observe that in the limit $T \rightarrow 0$, the 
thermal jamming transition density where the system becomes effectively non-ergodic is $\phi_{\textnormal{G}} = 0.55\pm 0.01$ for monodisperse spheres. To give a few examples from literature for comparison, for an experiment on colloidal suspensions with a small polydispersity $0.05$ a glass transition packing fraction $\phi_{\textnormal{G}}\approx 0.56$ is reported \cite{PuseyandMegen1}, while another experiment with a larger polydispersity indicates $\phi_{\textnormal{G}} \approx 0.58$ \cite{MasonandWeitz}. By fitting the power law divergence of the relaxation time predicted by mode-coupling theory \cite{GoetzeandSjoegren} to data of colloidal experiments $\phi_{\textnormal{G}}$ is expected in the range $0.571$ to $0.595$ \cite{vanMegen,Brambillaetal}. In computer simulations the dynamics was studied even beyond this prediction up to $\phi \approx 0.6$ \cite{Brambillaetal}. Numerical studies with soft repulsive harmonic spheres predict $\phi_{\textnormal{G}}$ at zero temperature limit at $0.637$ \cite{BerthierandWitten,BerthierandWitten1} or in the zero shear stress limit at $0.59$ \cite{IkedaBerthierSollich}. For emulsions, $0.589$ is reported for experiments \cite{jstat} and $0.591$ for simulations \cite{jstat}. 
Most of these differences probably are due to differences of the systems (e.g., monodisperse vs. polydisperse systems), different methods of extrapolations, or due to difficulties to determine packing fractions with high accuracy in experiments. However, as we will show in the following, different transition packing fractions might also arise due to different starting configurations.

\begin{figure}
\centering
\includegraphics[width=\linewidth]{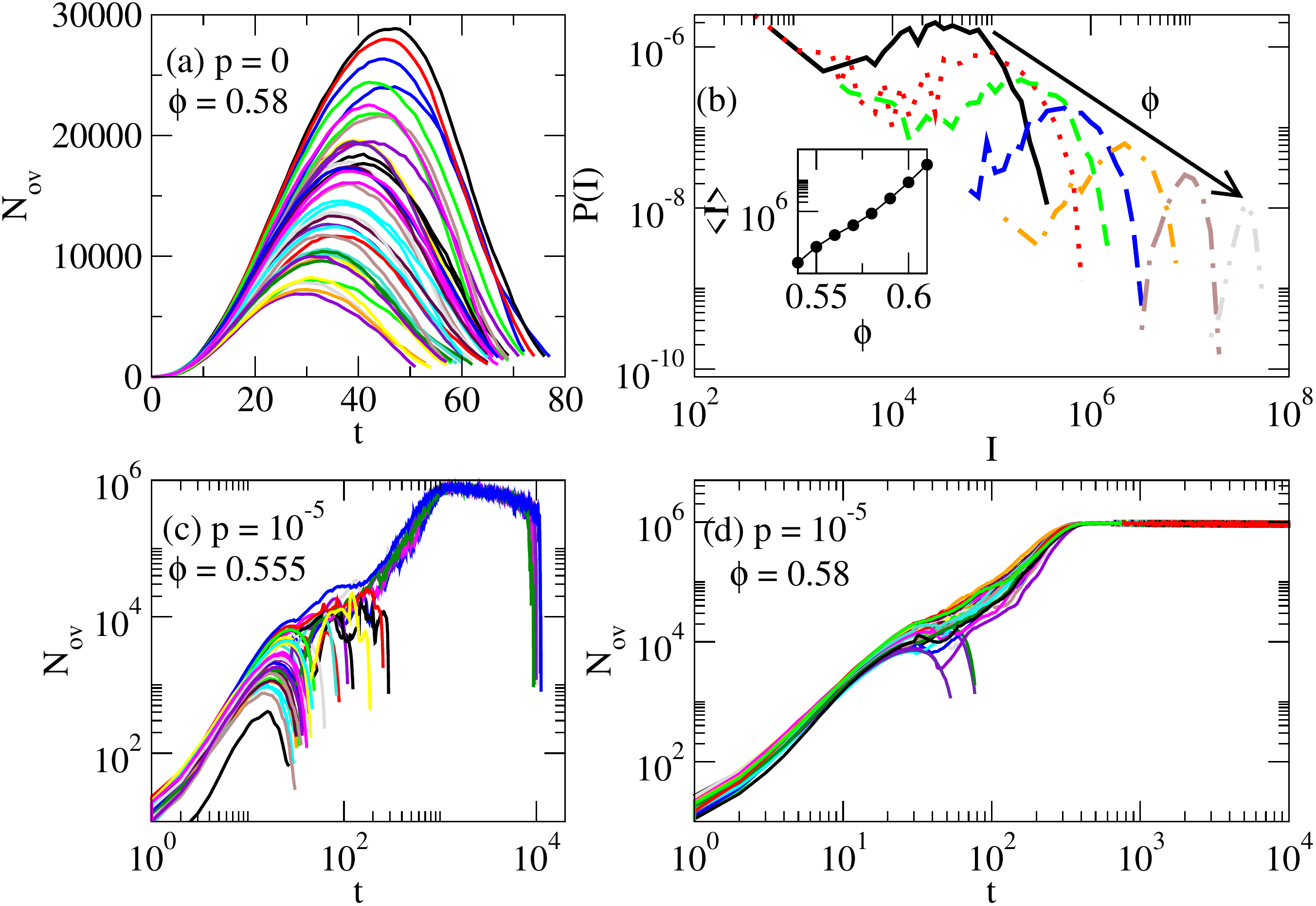}
\caption{
  \sffamily \small {\bf {\sffamily Starting configuration dependence of glass transition density.}}
  Analysis of how thermal moves affect systems that are athermally unjammed, i.e., in a local energy minimum determined with the athermal jamming protocol. In (a,b) we show how such a system relaxes if one particle 
crosses a barrier
In (c,d) one particle is moved 
over a barrier and then the thermal jamming protocol is switched on with $p=10^{-5}$. In (a,c,d) the number of overlapping particles as a function of time after the random move are shown. The different lines represent different realizations of the same protocol. In order to obtain (b), the area $I$ under the curves in (a) are determined. $I$ is a measure for how many events with overlapping particles occur during the course of time. The probabilities $P(I)$ how often a certain $I$ is realized is plotted in (b) for packing fractions $\phi=0.55$ (black line), $0.56$, $0.57$, $0.58$, $0.59$, $0.6$, and $0.61$ (dash-dotted gray line). The inset of (b) shows the mean values of these distributions as a function of packing fraction $\phi$.}
\label{fig:fig6}
\end{figure}

Instead of the random initial configurations that we have used so far, for the results shown in Fig.~\ref{fig:fig6} we employ athermally unjammed starting configurations that are obtained by the deterministic minimization protocol.
Note that we employ the usual athermal jamming protocol and not the modification used in \cite{corwin}, i.e., particles are not kept in contact at the end of the minimization. Furthermore, in the following only overlapping but not touching particles are considered for the analysis such that for our athermally unjammed starting configurations there is no spatially percolation for $\phi<\phi_{\textnormal{J}}$ while the thermally jammed configurations for small $p$ and $\phi>\phi_{\textnormal{G}}$ possess spatially percolated clusters as shown in the previous subsection.

In Fig.~\ref{fig:fig6}(a) we show the number of overlapping particles that we observe 
in case of an athermally 
unjammed starting configuration and for the athermal protocol after a single particle is moved over a barrier for $\phi=0.58$.
The different curves correspond to different realizations. Note that the system always ends up in an unjammed configuration and the number of overlapping particles always remains finite (and much smaller than the system size). In order to understand how further 
crossings of barriers
would affect such a system (e.g., for $p>0$) we determine how many overlapping particles exist integrated over time. The value $I$ of this integral indicates the total number of overlapping events. In Fig.~\ref{fig:fig6}(b) the probability distribution $P(I)$ of how often a given value $I$ occurs is plotted for various packing fractions ranging from $0.55$ to $0.61$. In none of the cases an infinite $I$ was observed. However, when increasing the packing fraction the $P(I)$ are shifted to larger $I$ and they become narrower. In the inset of Fig.~\ref{fig:fig6}(b) the mean values of $I$ are shown as a function of the packing fraction. Interestingly there is an exponential increase. 

Next, we want to find out whether we obtain a 
thermally jammed state if the thermal protocol with a $p>0$ is switched on. Fig.~\ref{fig:fig6}(c) shows the number overlaps as function of time after the 
initial barrier crossing
for various realizations for $p=10^{-5}$ and $\phi=0.555$ where in case of a random initial configuration a thermally jammed state occurs. However, if started from an athermally unjammed configuration the system is not in a 
thermally jammed
because for all realizations we find that the system relaxes into an unjammed configuration. Only at a larger density, e.g., for $\phi=0.58$ as plotted in Figs.~\ref{fig:fig6}(d), most realizations end up in a jammed state. Therefore, in case of an  athermally unjammed initial configuration and for $p=10^{-5}$ a 
thermal jamming transition
can be observed which occurs above $\phi=0.555$ but still well below $\phi_{\textnormal{J}}$. Note that we expect this apparent 
ergodicity breaking
transition to depend on $p$ even in the limit $p\rightarrow 0$ because there is no underlying spatial percolation transition. However, the dependence on $p$ will be weak (probably logarithmic) because a small increase in packing fraction results in an exponential increase in overlapping events as shown above. 
Finally, we want to point out that in order to study the hard-sphere limit it might be more natural to start with the non-overlapping configurations as considered in this subsection instead of random configurations that were employed in the previous sections and that can contain large overlaps. The behavior observed for small overlaps then in principle can be mapped onto hard sphere system \cite{Xuetal,HaxtonSchmiedebergandLiu}.

\section*{Discussion}

By adding the rare possibility to cross energy barriers to the protocol that previously was employed to study the athermal jamming transition, we obtain a powerful new method that allows the direct investigation of jamming at small but non-zero temperatures. By employing this method we determined whether the system can access the ground state or not. In the later case the system effectively is non-ergodic for small temperatures. Therefore, the observed transition is an effective ergodicity breaking transition.

We find that the ergodicity breaking transition is a directed percolation transition in time. The transition occurs at much smaller packing fractions than the athermal jamming transition and the transition density in the limit of small temperature does not depend on the temperature but is is given by the spatial percolation of particles in contact.

The specific value of the transition depends on the initial conditions. As we explained in our discussion of different starting configurations, past experiments and simulations used various protocols which might be one reason why different transition densities for the glass transition have been reported. Furthermore, as we explain in Supplementary Note 2, in too small systems the apparent ergodic to non-ergodic transition might be observed at a packing fraction that is larger than in our large system. 
However, for given initial conditions and large enough systems, our method can be used to directly predict the density of the ergodic to non-ergodic transition. For example, in case of a a very fast quenches from infinite to small final temperatures we find a transition at a packing fraction of 0.55 exactly as Morse and Corwin in \cite{corwin}.

Our protocol is constructed such that we can easily find out whether a ground state can be reached or not. Note that Brownian dynamics simulations or local Monte-Carlo simulations are superior for simulating fluctuations within a valley of the landscape or for producing an enable sampling of visited configurations. However, they are not superior in minimization and they all allow for the (maybe rare) crossing of barriers in case of non-zero temperature. As a consequence, Brownian dynamics or local Monte-Carlo simulations cannot reach a ground state that is not even accessible with our optimized approach. Note that while ensemble sampling with our method might require the use of additional weight factors, the power laws describing the critical behavior as well as the power law behavior for $g(r)$ shown in Supplementary Note 3 would not change if the resulting configurations had to be weighted differently in an ensemble sampling.

In recent years, there have been significant advances in studying and characterizing non-ergodic systems including those associated with anomalous diffusion (for reviews see, e.g., \cite{nonerg1,nonerg2}). Note that in case of the dynamical soft sphere glass transition that we study there might be rare random rearrangements even in the effectively non-ergodic glass phase. Due to these rearrangements the system is diffusive in the long-time limit. However, the diffusive motion occurs on a timescale that is longer than the timescale of observation and therefore the system is termed an effectively non-ergodic glass. In the language of stochastic dynamics such systems are termed weakly non-ergodic \cite{nonerg2}. With our approach in case of a small but non-zero probability $p$ for energy barrier crossings, the rearrangements are directly associated with such rare barrier crossing events. The timescale of the rearrangements is given by $1/p$ in units of minimization steps. The relaxation time typically is a multiple of this timescale. As a consequence the Deborah number $De$ that is defined as ratio of the timescale of relaxation and the timescale of observation $T$ \cite{deborah} has to be $De> 1/(p T)$, i.e., $De\rightarrow \infty$ for $p\rightarrow 0$. For example, in our case for $T\sim 10^5$ and $p=10^{-6}$ one finds $De>10$. Since the system is ergodic and diffusive in the long time limit, asymptotically the time average and the ensemble average of the squared displacement are the same. A corresponding weakly non-ergodic system is given by a Brownian particle that moves in a random, bounded external potential. For such a system the relation between time and ensemble average at finite times has been extensively studied in simulations as well as experiments \cite{randpot1,randpot2}.

We are confident that the new method can be employed to obtain more insights into the physics of glassy states, e.g., in order to study the properties of the modes in a thermally jammed but athermally unjammed state point (cf. \cite{Wangetal}), to determine the connection to rigidity percolation \cite{harowellpnas} and contact percolation \cite{ShenOHernandShattuck}, to understand a possible Gardner transition \cite{Gardner,GrossKanterandSompolinsky,Charbonneauetal}, to explore the properties of the basins in the energy landscape which can be used to determine the entropy of the system \cite{XuFrenkelandLiu,AsenjoPaillussonandFrenkel}, or extend the test of Edwards' approach to the statistics of granulates \cite{Edwards89,Edwards90} that recently has been tested for unjammed configurations \cite{Martinianietal}. Furthermore, it would be interesting to combine our method with alternative packing protocols \cite{torquato}, or to study the influence of shearing as in \cite{TjhungandBerthier}, to explore the glass transition for active particle \cite{henkes,berthieractive}, and to learn more about gelation in case of particles with short ranged attraction where recently a connection to spatial directed percolation has been reported \cite{Kohletal}.

\section*{Methods}
In case of athermal jamming we start with a random
configuration of spheres with diameter $\sigma$ in three dimensions in a cubic box with side length $L$ and periodic boundary conditions. We instantaneously quench the system to zero temperature by minimizing the total
energy. The interaction energy is given by the finite-ranged repulsive
harmonic pair potential, which is $V(r) = \epsilon(1 - r/\sigma)^2$ for
particle distances $r<\sigma$ and zero otherwise. The prefactor
$\epsilon$ sets the energy scale. The packing fraction is given by $\phi=\pi\sigma^3N/(6L^3)$. The energy landscape depends on the position of all particles and is the sum over all pair interaction energies. For minimization we use the conjugate
gradient method of the simulation package LAMMPS \cite{lammps}. The
number of minimization steps is denoted by $t$. The minimization is
stopped either if the number of overlapping particles as a function of
$t$ fluctuates around a plateau value (denoting a jammed state) or if
the energy per particle is $10^{-16}\epsilon$ or less (denoting an
unjammed state). Accordingly, we consider two particle to overlap if
$\sigma-r>10^{-7}\sigma$. We have checked that this energy cut-off and
the precision of the overlap definition do not have an influence on our
results.

In order to study jamming at a non-zero temperature we employ the same system but in addition
particles that are still part of the process, i.e., that are
either overlapping or touching (up to our precision,
$r-\sigma<10^{-7}\sigma$) are selected in each step with a probability
$p$. For the selected particles a random spatial direction is chosen ant the particles are displaced in that direction until
they reach a minimum or maximum of the total energy determined along the line in that direction. In the latter case
particles are set slightly behind the maximum such the energy barrier is
crossed. The LAMMPS software is modified to adapt it for our modified
protocol. Especially, the search direction to the gradient direction
during the relaxation are reset for the randomly displaced particles.
After selected particles have been displaced, all particles are moved
according to the minimization protocol in the same step. The stopping
criterion for the modified protocol is the same as for the athermal
jamming protocol.

In case of small $p$ large systems have to be considered in order to be
sure that the results are not affected by system size effects. For
example, for $p=10^{-6}$ close to the transition we employ systems with
up to $N=10^7$ particles. A detailed analysis of system size effects is
presented in Supplementary Note 2. We have checked
that no crystallization occurs by analyzing the bond orientational order
parameter $Q_{6}$ as described in \cite{Steinhardtetal}. Furthermore, we
have tested different implementations for the minimization as well as for the random steps (see Suplementary Note 1). Such details
affect how $p$ is defined, but not the transition density in the limit of
small $p$ or the type of the transition anywhere.

{\bf Data availability.} The data shown in this paper or the supplementary notes are available from the authors.

\newpage
\section*{Supplementary Notes on ``Ergodicity breaking transition in a glassy soft sphere system at small but non-zero temperatures''}
{\bf Moumita Maiti$^1$ and Michael Schmiedeberg$^{1,*}$}\\
$^1$Institut f{\"u}r Theoretische Physik 1, Friedrich-Alexander-Universit{\"a}t Erlangen-N{\"u}rnberg, Staudtstr. 7, 91058 Erlangen, Germany\\
$^*$michael.schmiedeberg@fau.de

\subsection*{Supplementary Note 1:\\Modifications of the protocol}
\begin{figure}
\begin{center}
  \includegraphics[width=1.\linewidth]{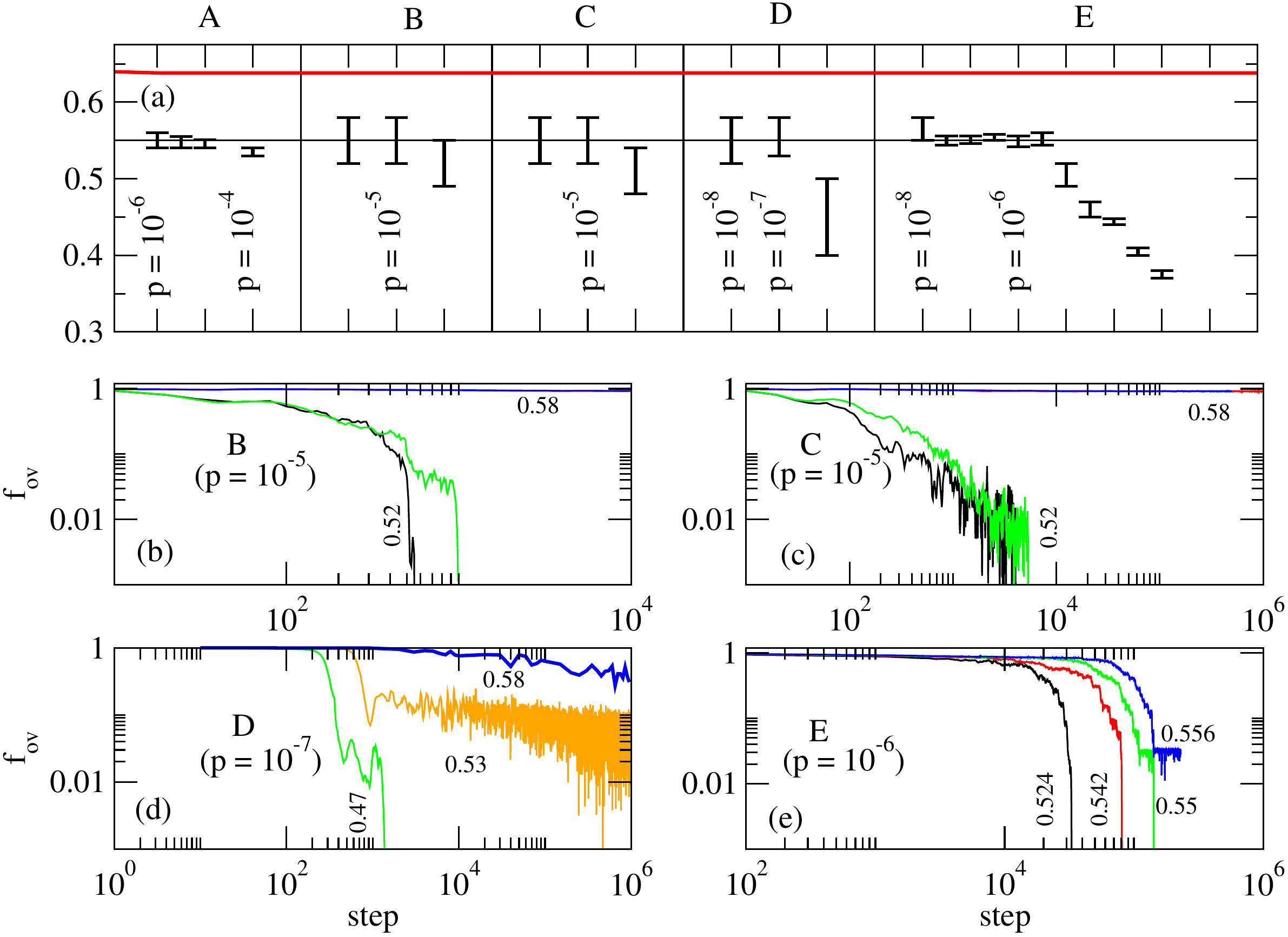}
  \end{center}
\caption{
  { {\bf Modifications of the protocol.}}
  (a) Ranges of the transition packing fraction for various models with different protocols. The models are labeled with the capital letters displayed on top of the figure and explained in the main text. The black horizontal line marks the packing fraction 0.55, the red horizontal line denotes the packing fraction of athermal jamming. (b-e) Fraction of overlapping particles $f_{\textnormal{ov}}$ as a function of step number $t$ for some of the cases for the models shown in (a). The respective models are marked by the capital letters in the figures. The numbers denote the packing fractions. The employed system sizes are the following: (b,c) $N=10^5$ (black and red lines) and $N=10^6$ (green and blue lines), (d) $N=10^6$, (e) average over $25$ runs with $N=25000$ particles.
}
\label{fig:suppfig1}
\end{figure}
In this section we present different models, where we vary the protocol that is employed to access the ergodicity breaking transition in the limit of rare barrier crossing events. In Fig. \ref{fig:suppfig1}(a) we summarize the ranges that are obtained for the transition packing fraction. The curves for the fraction of overlaps as a function of step number for cases close to this transition are displayed in Fig. \ref{fig:suppfig1}(b,c,d,e). The employed models are
\begin{itemize}
\item Model A: The model that is used in the main text and that is explained in detail therein.
\item Model B: In this variation of model A we only allow overlapping particles to take part in the protocol. Therefore, touching particles can no longer be randomly displaced.
\item Model C: Variation of model A, where in a random step particles are only displaced if in the chosen direction there is a crossing of an energy barrier. Otherwise the particles is not moved. Therefore, minimization only takes place during minimization steps.
\item Model D: The minimization is done according to the steepest descent algorithm. Otherwise the protocol of model A is used. Note that minimization using the steepest descend algorithm usually is less efficient than the conjugate gradient algorithm that is used in models A to C. As a consequence, the relaxation just below the transition is very slow and noisy as can be seen for the orange curve for $\phi=0.53$ in Fig. \ref{fig:suppfig1}(d). As a consequence, in order to differentiate between relaxing and non relaxing systems close to the transition, we also analyzed the structure. In case of a persisting percolated cluster we are sure to be in the non-ergodic phase, while below the transition the system disassociates into distinct non-percolating clusters of overlapping particles that slowly shrink.
\item Model E: In this model completely different protocols for minimization and random steps are used. The minimization steps corresponds to a deterministic athermal jamming step as used in \cite{MilzandSchmiedeberg}. In such a step two overlapping particles are deterministically displaced along the line given by the two center points of the particles until the particles are no longer overlapping but only touching. In random steps that occur with probability $p$ the touching particles in addition are rotated around the center of mass by an random angle. More details are given in \cite{MilzandSchmiedeberg}, where we studied the individual protocols but not their mixing. Note that the minimization in this model is very ineffective such that no large systems can be studied within a reasonable computation time. For the shown results, we averaged over 25 runs with $N=25000$ particles. As a consequence, the drop after a plateau has been reached for packing fraction 0.55 in Fig. \ref{fig:suppfig1}(e) probably is artifact due to the small system size. Therefore, in principle it is possible that the error bars shown in Fig. \ref{fig:suppfig1}(a) for model E differ from error bars obtained in larger systems. However, up to the system size that we could consider, even in this completely different model the plateaus start to develop at the same packing fraction as for the other models.
\end{itemize}

Concerning finite size effects, we want to point out that slower minimization protocols might require to go to smaller values of $p$ in order to reach the limit where the transition packing fraction no longer depends on $p$. As a consequence, such protocols might require larger system sizes to obtain a comparable accuracy for the transition packing fraction. In general, the width of the range for the transition packing fraction is mainly limited due to finite size effects. In model A we performed an extensive finite size analysis (cf. Supplementary Note 2) and as a consequence can narrow the range significantly. For models B and C we checked two different system size to show that the given range for the transition packing fraction is not affected by the system size. In case of model E we have to point out that significant finite size effects are still present. Especially the result for $p=10^{-8}$ is probably strongly affected by the system size.

In summary, in the limit of rare barrier crossing events, all models lead to a range for the transition packing fraction that is in agreement with the transition packing fraction determined by model A in the main text. 

\subsubsection*{Supplementary Note 2:\\Analysis of system size effects}

\begin{figure}
\begin{center}
\includegraphics[width=0.8\textwidth]{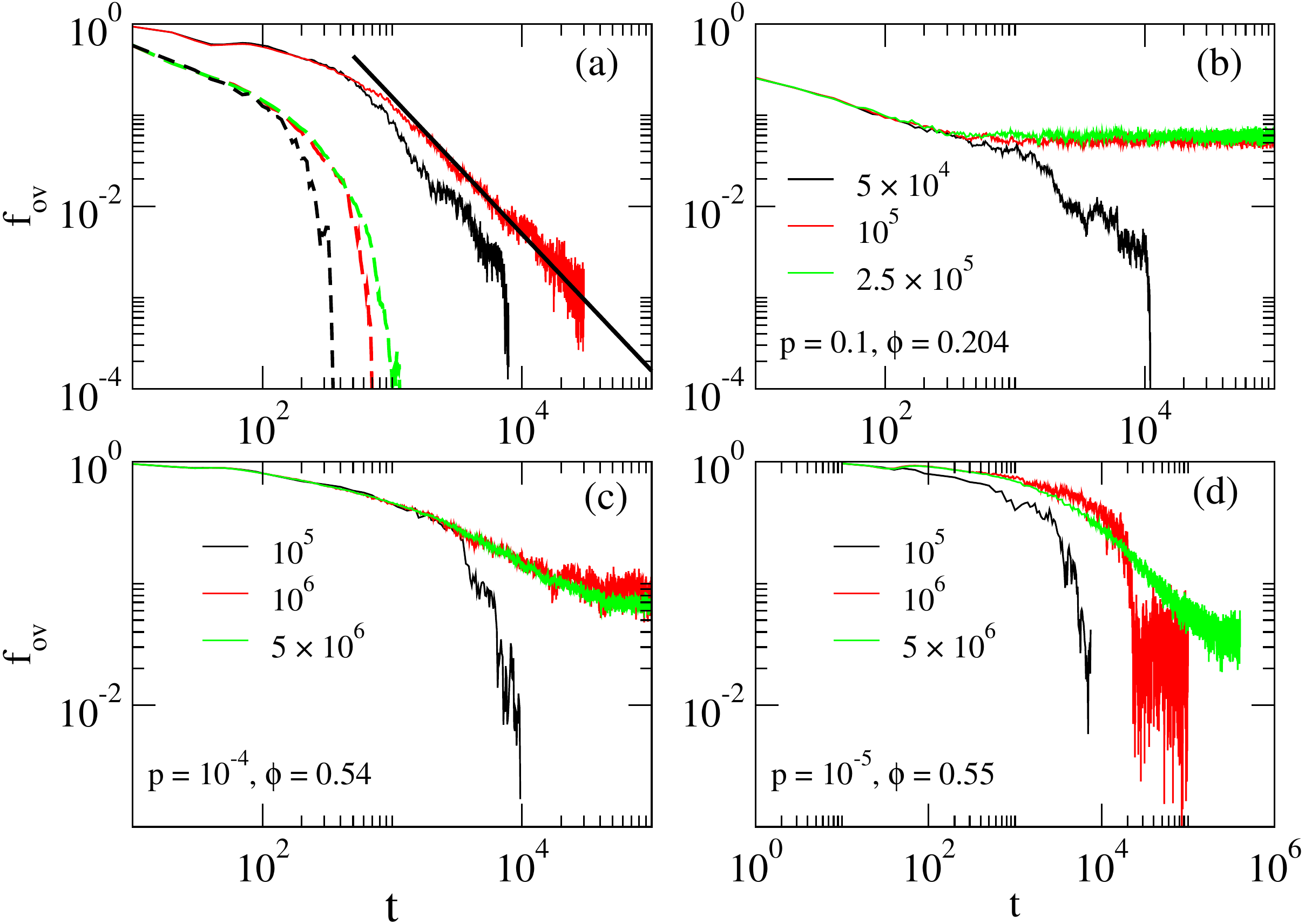}
\end{center}
	\caption{
		{ {\bf System size dependence.}
		Analysis of system size effects close to the thermal jamming transition for the relaxation process given by the fraction of overlapping particles $f_{\textnormal{ov}}(t)=N_{\textnormal{ov}}(t)/N$ as a function of step number $t$. (a) Typical finite size effects below the transition packing fraction for $p = 0.01$ (dashed lines) and for $p = 10^{-5}$ (solid lines). For $p = 0.01$ the packing fraction $\phi = 0.388$ just below the transition $\phi_{c} = 0.3998$ (as determined by a fit to the critical power law) is used for three system sizes $N = 10^{4}$ (black), $10^5$ (red), and $N = 5 \times 10^{5}$ (green). For $p = 10^{-5}$ we show data for $\phi = 0.52$ (which is below the transition at $\phi_{c} = 0.553$) for $N =10^{6}$ (black) and $N = 10^{7}$ (red). The black line indicates a power law with an exponent $-1.5$ shown for comparison. (b,c) Typical finite size effects above the transition packing fraction: (b) $p = 0.1$ and $\phi = 0.204$ (transition at $\phi_{\textnormal{c}} = 0.199$ determined by a fit to the critical power law) and (c) $p = 10^{-4}$ and $\phi = 0.54$ ($\phi_{c} = 0.5398$). (d) Finite size effects close to the transition for $p = 10^{-5}$ and $\phi = 0.55$.
		\label{fig:suppfig2}}}
\end{figure}

In the following we show how the relaxation process might be affected by finite size effects. Furthermore, at the end of this section we explain how we choose our system sizes such that the transitions that we determine do not depend on these sizes.

Fig.~\ref{fig:suppfig2}(a) demonstrates that below the transition packing fraction smaller systems relax sooner than larger systems. For the small probability $p = 10^{-5}$ we point out to another interesting observation: We find that the relaxation obeys an unexpected power law with an exponent $-1.5$. Note that this power law only occurs below the transition and is not related to the critical power law decay $t^{\alpha}$ with $\alpha = -0.732$ which occurs directly at the transition. 

For the relaxation above the transition, we depict different system sizes for $p = 0.1$ in Fig.~\ref{fig:suppfig2}(b) and for  $p = 10^{-4}$ in Fig.~\ref{fig:suppfig2}(c). For small systems the fraction of overlapping particles might suddenly decay to zero instead of approaching the plateau value that should occur in the long-time limit. Therefore, in small systems it is easier to find the unjammed configuration while in large systems the zero energy states are not reached due to the larger configuration space. As a consequence of the system size analysis one can conclude, that if too small systems were considered, one would observe a ergodicity breaking only at larger densities. For example, in the limit $p\rightarrow 0$ the apparent packing fraction of the ergodic to non-ergodic 
transition would be much larger than $0.55$ if systems with $10^3$ to $10^4$ particles were considered as it is the case in many simulations of glassy dynamics. 

In Fig.~\ref{fig:suppfig2}(d) we also show a case for $p = 10^{-5}$ where an unexpected jump at intermediate system sizes occurs (red curve for $N=10^6$), where the relaxation first seems to decay rapidly but then fluctuates around a non-zero plateau value. Such a behavior is only observed in a few cases at small $p$ and close to the transition.

In order to determine the thermal jamming transition independent of the system size we use the following recipe: The system size for a given $p$ at first is chosen from the criteria $Np = 1$ such that at each step one random move is performed, then we increase $N$ until for the curves that are closest to the transition we do not observe any significant change for two different system sizes. 

The system sizes that we determined by this method and that we actually used for the curves shown in Fig. 2 are $N=50000$ for $p=1$, $N=10^5$ for $p=0.1$, $N=10^6$ for $p=10^{-3}$ and $p=10^{-4}$ as well as for $p=10^{-5}$ below the transition, $N=5\times 10^6$ for $p=10^{-5}$ above the transition, and $N=8\times 10^6$ for $p=5\times 10^{-6}$. In order to further narrow the ranges in which the transition occurs, we sometimes employ even larger systems in Fig. 3 and Fig. 4 with up to $N=10^7$ particles, e.g., in case of $p= 10^{-6}$.

Note that the dependence on system size is most prominent close to the transition. As a consequence the system size in principle might limit how close we can get to the transition without being affected by finite size effects. For the resolution of $\phi$ that we used in our article this usually is not a problem. Only for $p> 10^{-5}$ we were not able to further narrow the range of the transition packing fraction, e.g., in Fig. 4 of the main article we cannot report any plateau values $f_{\textnormal{ov}}(t\rightarrow \infty)$ close to $0$ for $p> 10^{-5}$ because only data that has proven to be independent of the system size is shown. Furthermore, for $p\geq 10^{-5}$ we are not able to determine values of $\tau$ that do not depend on the choice of the fitting range or how the data points of the relaxation curve are spaced. The reason is that for small $p$ we are not able to get close enough to the transition such that the range of the power law behavior that is assumed for the fitting (see also Fig.~\ref{fig:suppfig3}(a)) is too small in order to obtain a reliable fit. Therefore, there are no $\tau$-values for $p\geq 10^{-5}$ in Fig. 4(b) but only values for $f_{\textnormal{ov}}(t\rightarrow \infty)$ in Fig. 4(a) which is still independent of the way of fitting. If one wanted to obtain values of $\tau$ for smaller $p$ one would need larger systems closer to the transition such that the power law behavior would be sufficiently extended.

\subsubsection*{Supplementary Note 3:\\ Close to the thermal jamming transition: Relaxation curves and pair correlation functions}

\begin{figure}
\begin{center}
\includegraphics[width=0.8\textwidth]{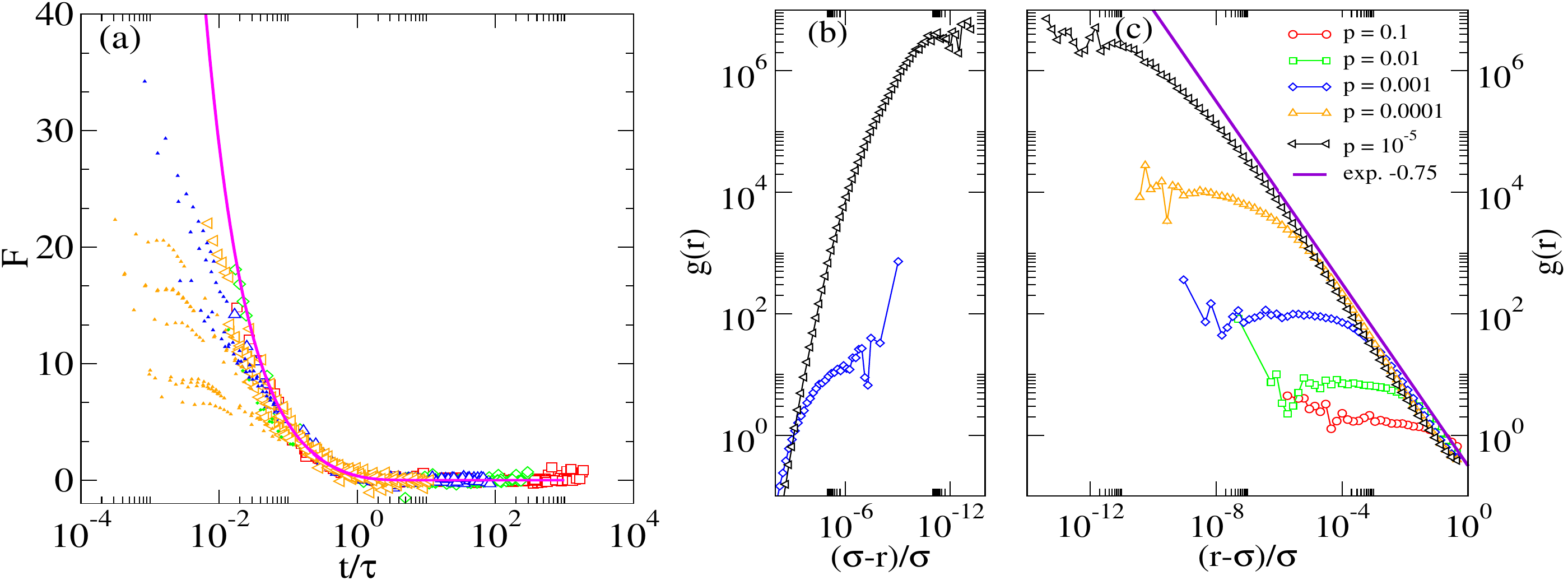}
\end{center}
\caption{
		{ {\bf Relaxation curve and pair correlation function.}
		(a) Collapse of relaxation functions to demonstrate the agreement with Eq.(2). $F= \tau^{0.732}(f_{\textnormal{ov}} - f_{\textnormal{ov}}(t \rightarrow \infty))/A$ as a function of $\frac{t}{\tau}$ is plotted for different $p$ for cases close to the transition. The colors are the same as in (c). All curves collapse onto $(t/\tau)^{-0.732}\exp\left(-t/\tau\right)$ (magenta line) as expected from Eq.(2). (b) Log-log-plot of the pair correlation function $g(r)$ shown for distances $\sigma-r$, i.e., for small overlaps. (c) Log-log-plot of the pair correlation function $g(r)$ shown for distances $r-\sigma$, i.e., for the cases where particles are close to contact but do not touch. The curves (b) and (c) are shown for structures obtained at packing fractions just above the thermal jamming transition. Before the first minimum in (c), all curves decay like a power law with exponent $-0.75$ (magenta line).  
		\label{fig:suppfig3}}}
\end{figure}

The fraction of overlapping particles $f_{\textnormal{ov}}(t)=N_{\textnormal{ov}}(t)/N$ as a function of step number $t$ decreases like a power law $t^{\alpha}$ close to the thermal jamming transition. Close to the transition, the relaxation curves can be fitted with the function given in Eq.(2) where we assumed $\alpha =\frac{\beta}{\nu}= -0.732$ as for directed percolation. The quality of this choice of functional behavior is demonstrated in Fig.~\ref{fig:suppfig3}(a) where we collapse our data for the relaxation curves with different $p$ and for $\phi$ just above the transition by plotting $F = \tau^{0.732}(f_{\textnormal{ov}} - f_{\textnormal{ov}}(t \rightarrow \infty))/A$ as a function of $\frac{t}{\tau}$. All curves are in good agreement with $(t/\tau)^{-0.732}\exp\left(-t/\tau\right)$ (magenta line) as expected from Eq.(2).

In Figs.~\ref{fig:suppfig3}(b) and (c) we plot the pair correlation function $g(r)$ for thermally jammed structures with $\phi$ just above the thermal jamming transition and for various $p$. In case of the non-overlapping part shown in Fig.~\ref{fig:suppfig3}(c), between the contact distance and the first minima of $g(r)$ we observe a power law decay with an exponent $-0.75$. The exponent is robust for all $p$. For $p\rightarrow 0$ the power law extends closer to the contact point. Note that for athermal jamming a power law of $g(r)$ is also observed \cite{jammingpowerlaw}. However, in case of athermal jamming the exponent is $-0.5$ \cite{jammingpowerlaw} which significantly differs from the exponent $-0.75$ that we observe here for all values of $p$. Therefore not only the nature of the thermal jamming transition is fundamentally different from the athermal jamming transition as demonstrated in the main text but also for the structural properties it makes a huge difference whether a system is athermally jammed or thermally jammed.

\section*{Acknowledgements}

The work was supported by the German Research Foundation (DFG) by grant Schm 2657/3. We thank Lars Milz for helpful discussions.

\section*{Author contributions statement}
M.M. carried out the simulations and M.S. designed the research and the model system. Both authors analyzed the results and wrote the article.

\section*{Additional information}

\textbf{Accession codes} (where applicable).\\
\textbf{Competing financial interests:} The authors declare no competing financial interests.


\newpage
\begin{thebibliography}{10}
\bibitem{Tabor}
Tabor, D. Gases, Liquids, and Solids (and Other States of Matter) (Cambridge University Press, Cambridge). (1991).

\bibitem{Angell}
Angell, C.~A. Formation of glasses from liquids and biopolymers. {\it Science} {\bf 267}, 1924--1935 (1995).

\bibitem{Debenedetti}
Debenedetti, P.~G. and Stillinger, F.~H. Supercooled liquids and the glass transition. {\it Nature (London)} {\bf 410}, 259--267 (2001).

\bibitem{Richardetal}
Richard, P., Nicodemi, M., Delannay, R., Ribi{\'e}re, P., and Bideau, D. Slow relaxation and compaction of granular systems. {\it Nature Materials} {\bf 4}, 121--128 (2005).

\bibitem{HunterandWeeks}
Hunter, G.~L. and Weeks, E.~R. The physics of the colloidal glass transition.
{\it Rep. Prog. Phys.} {\bf 75}, 066501 (2012).

\bibitem{PuseyandMegen}
Pusey, P.~N. and van Megen, W. Phase behaviour of concentrated suspensions of nearly hard colloidal spheres. {\it Nature} {\bf 320}, 340--342 (1986).

\bibitem{PuseyandMegen1}
Pusey, P.~N. and van Megen, W. Observation of a glass transition in suspensions of spherical colloidal particles. {\it Phys. Rev. Lett.} {\bf 59}, 2083--2086 (1987).

\bibitem{kegel}
Kegel, W.~K. and van Blaaderen, A.
Direct Observation of Dynamical Heterogeneities in Colloidal Hard-Sphere Suspensions. {\it Science} {\bf 287}, 290--293 (2000).

\bibitem{weeks2000}
Weeks, E.~R., Crocker, J.~C., Levitt, A.~C., Schofield, A., and Weitz, D.~A.
Three-Dimensional Direct Imaging of Structural Relaxation Near the Colloidal Glass Transition. {\it Science} {\bf 287}, 627--631 (2000).

\bibitem{WoodcockandAngell}
Woodcock, L.~V.and Angell, C.~A. Diffusivity of the hard sphere model in the region of fluid metastability. {\it Phys. Rev. Lett.} {\bf 47}, 1129--1132 (1981).

\bibitem{Speedy}
Speedy, R.~J. The hard sphere glass transition. {\it Mol. Phys.} {\bf 95}, 169--178 (1998).

\bibitem{Heuer}
Doliwa, B. and Heuer, A. Cage Effect, Local Anisotropies, and Dynamic Heterogeneities at the Glass Transition: A Computer Study of Hard Spheres. {\it Phys. Rev. Lett.} {\bf 80}, 4915--4918 (1998).

\bibitem{Sear}
Sear, R.~P. Molecular dynamics of a dense fluid of polydisperse hard spheres.
{\it J. Chem. Phys.} {\bf 113}, 4732--4739 (2000).

\bibitem{Dyre}
Dyre, J.~C. Colloquium: The glass transition and elastic models of glass-forming liquids. {\it Rev. Mod. Phys.} {\bf 78}, 953 (2006).

\bibitem{BerthierandBiroli}
Berthier, L. and Biroli, G. Theoretical perspective on the glass transition and amorphous materials. {\it Rev. Mod. Phys.} {\bf 83}, 587--645 (2011).


\bibitem{LubachevskyandStillinger}
Lubachevsky, B.~D. and Stillinger, F.~H. Geometric properties of random disk packings. {\it J. Stat. Phys.} {\bf 60}, 561--583 (1990).

\bibitem{LiuandNagel1}
Liu, A.~J. and Nagel, S.~R. The jamming transition and the marginally jammed solid. {\it Annu. Rev. Condens. Matter Phys.} {\bf 1}, 347--369 (2010).

\bibitem{OhernLangerLiuNagel}
O'Hern, C.~S., Langer, S.~A., Liu, A.~J., and Nagel, S.~R. Random packings of frictionless particles. {\it Phys. Rev. Lett.} {\bf 88}, 075507 (2002).

\bibitem{OhernSilbertLiuNagel}
O'Hern, C.~S., Silbert, L.~E., Liu, A.~J., and Nagel, S.~R. Jamming at zero temperature and zero applied stress: The epitome of disorder. {\it Phys. Rev. E} {\bf 68}, 011306 (2003).

\bibitem{Chaudhuryetal}
Chaudhury, P., Berthier, L., and Sastry, S. Jamming transitions in amorphous packings of frictionless spheres occur over a continuous range of volume fractions. {\it Phys. Rev. Lett.} {\bf 104}, 165701 (2010).

\bibitem{Bertrandetal}
  Bertrand, T. {\it et al.}
  Protocol dependence of the jamming transition. {\it Phys. Rev. E} {\bf 93}, 012901 (2016).

\bibitem{LiuandNagel}
Liu, A.~J. and Nagel, S.~R. Nonlinear dynamics:  Jamming is not just cool any more. {\it Nature} {\bf 396}, 21--22 (1998).
 
\bibitem{krzakalaandKurchan}
Krzakala, F. and Kurchan, J. Landscape analysis of constraint satisfaction problems. {\it Phys. Rev. E} {\bf 76}, 021122 (2007).

\bibitem{parisiandzamponi}
Parisi, G. and Zamponi, F. Mean-field theory of hard sphere glasses and jamming. {\it Rev. Mod. Phys.} {\bf 82}, 789--845 (2010).

\bibitem{JacquinBerthierandZamponi}
Jacquin, H., Berthier, L., and Zamponi, F. Microscopic mean-field theory of the jamming transition. {\it Phys. Rev. Lett.} {\bf 106}, 135702 (2011)

\bibitem{zhangetal}
  Zhang, Z. {\it et al.}  
  Thermal vestige of the zero-temperature jamming transition. {\it Nature} {\bf 459}, 230--233 (2009).

\bibitem{BerthierandWitten}
Berthier, L. and Witten, T.~A. Compressing nearly hard sphere fluids increases glass fragility. {\it Europhys. Lett.} {\bf 86}, 10001 (2009).

\bibitem{BerthierandWitten1}
Berthier, L. and Witten, T.~A. Glass transition of dense fluids of hard and compressible spheres. {\it Phys. Rev. E} {\bf 80}, 021502 (2009).

\bibitem{IkedaBerthierSollich}
Ikeda, A., Berthier, L., and Sollich, P. Unified study of glass and jamming rheology in soft particle systems. {\it Phys. Rev. Lett.} {\bf 109}, 018301 (2012).

\bibitem{Wangetal}
  Wang, X., Zheng, W., Wang, L., and Xu, N. Disordered solids without well-defined transverse phonons: the nature of hard-sphere glass. {\it Phys. Rev. Lett.} {\bf 114}, 035502 (2015).

\bibitem{corwin} Morse, P.~K. and Corwin, E~I. 2017 Echoes of the glass transition in athermal soft spheres. {\it Phys. Rev. Lett.} {\bf 119}, 118003 (2017).

\bibitem{charb11} Charbonneau, P., Ikeda, A., Parisi, G. and Zamponi F. Glass transition and random close packing above three dimensions. {\it Phys. Rev. Lett.} {\bf 107}, 185702 (2011).

\bibitem{buechner00} B\"uchner, S. and Heuer, A. Metastable states as a key to the dynamics of supercooled liquids. {\it Phys. Rev. Lett.} {\bf 84}, 2168 -2171 (2000).

\bibitem{denny03} Denny, R.~A., Reichman, D.~R. and Bouchaud, J.~P. Trap models and slow dynamics in supercooled liquids. {\it Phys. Rev. Lett.} {\bf 90}, 025503 (2003).

\bibitem{doliwa03} Doliwa, B. and Heuer, A. Energy barriers and activated dynamics in a supercooled Lennard-Jones liquid. {\it Phys. Rev. E} {\bf 67} 031506 (2003).

  \bibitem{Pine}
Pine, D.~J., Gollub, J.~P., Brady, J.~F., and Leshansky, A.~M. Chaos and threshold for irreversibility in sheared suspensions. {\it Nature} {\bf 438}, 997--1000 (2005).

\bibitem{Corte} Cort{\'e}, L., Chaikin, P.~M., Gollub, J.~P., and Pine, D.~J. Random organization in periodically driven systems. {\it Nature Phys.} {\bf 4}, 420--424 (2008).
  
\bibitem{Kramers} Hänggi, P., Talkner, P., and Borkovec, M. Reaction-rate theory: fifty years after Kramers {\it Rev. Mod. Phys.} {\bf 62}, 251 (1990).
  
\bibitem{MilzandSchmiedeberg}
Milz, L. and Schmiedeberg, M. Connecting the random organization transition and jamming within a unifying model system. {\it Phys. Rev. E} {\bf 88}, 062308 (2013).

\bibitem{Frenkel}
  Schrenk, K.~J. and Frenkel, D. Communication: Evidence for non-ergodicity in quieschent states of periodically sheared suspensions. {\it J. Chem. Phys.} {\bf 143}, 241103 (2015).
   
\bibitem{TjhungandBerthier}
Tjhung, E. and Berthier, L. Hyperuniform density fluctuations and diverging dynamic correlations in periodically driven colloidal suspensions. {\it Phys. Rev. Lett.} {\bf 114}, 148301 (2015).

\bibitem{ReichhardtandReichhardt}
Reichhardt, C. and Reichhardt, C.~J.~O. Random organization and plastic depinning. {\it Phys. Rev. Lett.} {\bf 103}, 168301 (2009).
 
\bibitem{Hinrichsen}
Hinrichsen, H. Nonequilibrium critical phenomena and phase transitions into absorbing states. {\it Adv. Phys.} {\bf 49}, 815--958 (2000).

\bibitem{MasonandWeitz}
Mason, T.~G. and Weitz, D.~A. Linear viscoelasticity of cooloidal hard sphere suspensions near the glass transition. {\it Phys. Rev. Lett.} {\bf 75}, 2770-2773 (1995).

\bibitem{GoetzeandSjoegren}
G{\"o}tze, W. and Sj{\"o}gren, L. Relaxation processes in supercooled liquids. {\it Rep. Prog. Phys.} {\bf 55}, 241--376 (1992).

\bibitem{vanMegen}
van Megen, W., Mortensen, T.~C., Williams, S.~R., and M{\"u}ller, J. Measurement of the self-intermediate scattering function of suspensions of hard spherical particles near the glass transition. {\it Phys. Rev. E} {\bf 58}, 6073-6085 (1998).

\bibitem{Brambillaetal}
  Brambilla, G. {\it et al.}
  Probing the equilibrium dynamics of colloidal hard spheres above the mode-coupling glass transition. {\it Phys. Rev. Lett.} {\bf 102}, 085703 (2009).
  
\bibitem{jstat}
  Zhang, C., Gnan, N., Mason, T.~G., Zaccarelli, E., and Scheffold, F. Dynamical and structural signatures of the glass transition in emulsions. {\it J. Stat. Mech.: Theory and Experiments} {\bf 2016}, 094003 (2016).

\bibitem{Xuetal}
Xu, N., Haxton, T.~K., Liu, A.~J., and Nagel, S.~R. Equivalence of glass transition and colloidal glass transition in the hard-sphere limit. {\it Phys. Rev. Lett.} {\bf 103}, 245701 (2009).

\bibitem{HaxtonSchmiedebergandLiu}
Haxton, T.~K., Schmiedeberg, M., and Liu, A.~J. Universal jamming phase diagram in the hard sphere limit. {\it Phys. Rev. E} {\bf 83}, 031503 (2011). 

\bibitem{nonerg1}
Metzler, R., Jeon, J.-H., Cherstvy, A.~G., and Barkai, E. Anomalous diffusion models and their properties: non-stationarity, non-ergodicity, and ageing at the centenary of single particle tracking. {\it Phys. Chem. Chem. Phys.} {\bf 16}, 24128 (2014).

\bibitem{nonerg2}
Meroz, Y. and Sokolov, I.~M. A toolbox for determining subdiffusive mechanisms. {\it Physics Reports} {\bf 573}, 1 (2015).

\bibitem{deborah}
Reiner, M. The Deborah Number. {\it Physics Today} {\bf 17}, 62 (1964).

\bibitem{randpot1}
Hanes, R.~D.~L., Schmiedeberg, M., and Egelhaaf, S.~U. Brownian particles on rough substrates: Relation between intermediate subdiffusion and asymptotic long-time diffusion. {\it Phys. Rev. E} {\bf 88}, 062133 (2013).

\bibitem{randpot2}
  Bewerunge, J. {\it et al.}
  Time- and ensemble-averages in evolving systems: the case of Brownian particles in random potentials. {\it Phys. Chem. Chem. Phys.} {\bf 18}, 18887 (2016).


\bibitem{harowellpnas}
  De Souza, V. and Harrowell, P. Rigidity percolation and the spatial heterogeneity of soft modes in disordered materials. {\it Proc. Natl. Acad. Sci. USA} {\bf 106}, 15136--15141 (2009).

  \bibitem{ShenOHernandShattuck}
    Shen, T., O'Hern, C.~S., and Shattuck, M.~D. The contact percolation transition in athermal particulate systems. {\it Phys. Rev. E} {\bf 85}, 011308 (2012).
  
\bibitem{Gardner}
Gardner, E. Spin glasses with p-spin interactions. {\it Nucl. Phys. B} {\bf 257}, 747-765 (1985).

\bibitem{GrossKanterandSompolinsky}
Gross, D.~J., Kanter, I., and Sompolinsky, H. Mean-field theory of the Potts glass.{\it Phys. Rev. Lett.} {\bf 55}, 304-307 (1985).

\bibitem{Charbonneauetal}
  Charbonneau, P. {\it et al.}
  Numerical detection of the Gardner transition in a mean-field glass former. {\it Phys. Rev. E} {\bf 92}, 012316 (2015).

\bibitem{XuFrenkelandLiu}
Xu, N., Frenkel, D., and Liu, A.~J. Direct determination of the size of basins of attraction of jammed solids. {\it Phys. Rev. Lett.} {\bf 106}, 245502 (2011).

\bibitem{AsenjoPaillussonandFrenkel}
Asenjo, D., Paillusson, F., and Frenkel, D. Numerical calculations of granular entropy. {\it Phys. Rev. Lett.} {\bf 112}, 098002 (2014).


\bibitem{Edwards89}
Edwards, S. and Oakeshott, R. Theory of powders. {\it Physica A} {\bf 157}, 1080--1090 (1989).

\bibitem{Edwards90}
Edwards, S. The flow of powders and of liquids of high viscosity. {\it J. Phys.: Condens. Matter} {\bf 2}, SA63--SA68 (1990).

\bibitem{Martinianietal}
  Martiniani, S., Schrenk, K,~J., Ramola, K., Chakraborty, B., and Frenkel, D. Are some packings more equal than others? A direct test of the Edwards conjecture. arXiv:1610.06328 (2016).

\bibitem{torquato} 
  S.~Torquato, ``Random Heterogeneous Materials: Microstructure and Macroscopic Properties,'' (Springer Science \& Business Media, New York, 2002).
  
\bibitem{henkes}
Henkes, S., Fily, Y., and Marchetti, M.~C. Active jamming: self-propelled soft particles at high density. {\it Phys. Rev. E} {\bf 84}, 040301 (2011).

\bibitem{berthieractive}
Berthier, L. and Kurchan, J. Non-equilibrium glass transitions in driven and active matter. {\it Nature Physics} {\bf 9}, 310–314 (2013).

\bibitem{Kohletal}
Kohl, M., Capellmann, R.~F., Laurati, M., Egelhaaf, S.~U., and Schmiedeberg, M. Directed percolation identified as equilibrium pre-transition towards non-equilibrium arrested get states. {\it Nature Communications} {\bf 7}, 11817 (2016).

\bibitem{lammps}
 Plimpton, S. Fast parallel algorithms for short-range molecular dynamics. {\it J. Comp. Phys.} {\bf 117}, 1--19 (1995).

\bibitem{Steinhardtetal}
  Steinhardt, P.~J., Nelson, D.~R., and Ronchetti, M. Bond-orientational order in liquids and glasses. {\it Phys. Rev. B} {\bf 28}, 784--805 (1983).

\bibitem{jammingpowerlaw} 
Silbert, L.~E., Liu, A.~J., and Nagel, S.~R. Structural signatures of the unjamming transition at zero temperature. {\it Phys. Rev. E} {\bf 73}, 041304 (2006).

  
\end{thebibliography}
\end{document}